\documentclass[]{modaastex63}

\usepackage{amsmath}
\usepackage{tabularx}
\usepackage{extarrows}
\usepackage{placeins}
\shorttitle{Determination of 1929 Asteroid Rotation Periods from WISE Data}
\shortauthors{Lam, Margot, Whittaker, Myhrvold}

\usepackage{hyperref}
\hypersetup{colorlinks=true,citecolor=blue,linkcolor=blue,urlcolor=blue,breaklinks=true}

\begin{document}

\title{Determination of 1929 Asteroid Rotation Periods from WISE Data}

\correspondingauthor{Jean-Luc Margot}
\email{jlm@epss.ucla.edu}

\author[0000-0002-4688-314X]{Adrian L.H. Lam}
\affiliation{Department of Physics and Astronomy, University of California, Los Angeles, CA 90095, USA}
\affiliation{Department of Electrical Engineering, University of California, Los Angeles, CA 90095, USA}

\author[0000-0001-9798-1797]{Jean-Luc Margot}
\affiliation{Department of Earth, Planetary, and Space Sciences, University of California, Los Angeles, CA 90095, USA}
\affiliation{Department of Physics and Astronomy, University of California, Los Angeles, CA 90095, USA}

\author[0000-0002-1518-7475]{Emily Whittaker}
\affiliation{Department of Earth, Planetary, and Space Sciences, University of California, Los Angeles, CA 90095, USA}

\author[0000-0003-3994-5143]{Nathan Myhrvold}
\affiliation{Intellectual Ventures, 3150 139th Ave SE, Bellevue, WA 98005, USA}

\begin{abstract}
We used 22 $\mu$m (W4) Wide-field Infrared Survey Explorer (WISE)
observations of 4420 asteroids to analyze lightcurves and determined
spin period estimates for 1929 asteroids.  We fit second-order Fourier
models at a large number of trial frequencies to the W4 data and
analyzed the resulting periodograms.  We initially excluded rotational
frequencies exceeding 7.57 rotations per day ($P < 3.17$~hr), which
are not sampled adequately by WISE, and periods that exceed twice the
WISE observation interval, which is typically 36 hr.  We found that
three solutions accurately capture the vast majority of the rotational
frequencies in our sample: the best-fit frequency and its mirrors
around 3.78 and 7.57 rotations per day.  By comparing our solutions to
a high-quality control group of 752 asteroid spin periods, we found
that one of our solutions is accurate (within 5\%) in 88\% of the
cases.  The best-fit, secondary, and tertiary solutions are accurate
in 55\%, 27\%, and 6\% of the cases, respectively.
\added{We also observed that suppression of aliased solutions was more effective with non-uniform
  sampling than with quasi-uniform sampling.}

\end{abstract}

\section{Introduction}
\label{sec-intro}

The rotational period of an asteroid is a physical property that is important in a wide range of planetary science and space exploration contexts.   A rotational period measurement is essential to the characterization of the rotational state~\citep[e.g.,][]{prav02}, which informs our understanding of an asteroid's interior and morphology~\citep[e.g.,][]{Sche15AIV}, dynamical evolution through the Yarkovsky and YORP effects~\citep[e.g.,][]{Vokr15AIV,gree20}, and the formation and evolution of binaries, triples, and pairs~\citep[e.g.,][]{Marg15AIV,Wals15AIV}.  Spin rate distributions place bounds on the dynamical and collisional evolutions of the main belt of asteroids~\citep[e.g.,][]{bott15AIV}, and therefore the characteristics of the near-Earth asteroid population, which governs the history of impact cratering in the inner solar system and affects planetary defense efforts.  Spin periods also provide useful initial conditions when modeling the shape~\citep[e.g.,][]{Ostr02,Benn15AIV,Dure15AIV} and thermophysical properties~\citep[e.g.,][]{Delb15AIV} of asteroids.
In these contexts, the availability of a small number of candidate spin periods is extremely valuable.  One can test the model with each trial period and promptly identify the correct period.  This fact motivated in part our inclusion of secondary spin period solutions in our results, in addition to our primary, best-fit period solutions.

Several approaches have been used to measure the rotational periods of asteroids with high precision, including Earth-based~\citep[e.g.,][]{prav02} or space-based~\citep[e.g.,][]{hora18} photometric observations or a combination of the two~\citep[e.g.,][]{dure18}, as well as Earth-based radar observations~\citep[e.g.,][]{ostr06,naid15dp}.  Optical lightcurve photometry based on wideband measurements of the sunlight reflected by the asteroid is the most common approach.
Here we use infrared lightcurves to determine asteroid spin periods.

During its six-month primary mission, the WISE spacecraft~\citep{wrig10} conducted a whole-sky infrared survey at four infrared bands (W1--4) centered at 3.4, 4.6, 12, and 22 $\mu$m.   All four detectors were simultaneously exposed, producing up to four independent photometric measurements \citep{wisedata}.  The high-quality, multi-band IR observations of $\sim$100,000 asteroids have been used to estimate asteroid diameters and albedos~\citep[e.g.,][]{main15AIV}.  Improved algorithms applied to a curated set of thousands of asteroids yielded refined estimates as well as estimates for asteroids not previously analyzed~\citep{myhr22}.

Both the cadence and length of an asteroid lightcurve observing campaign determine the parameters that can be reliably recovered from the observations.  Durations on the order of days/months/years, such as survey data from the Palomar Transient Factory (PTF) \citep{wasz15}, provide an opportunity to sample the object at different phase angles and to recover parameters of the phase function \citep{muin10}.  Densely sampled observations that span at least a complete rotational cycle provide the best opportunity to determine the spin period.  WISE observations of asteroids present a challenge for lightcurve analyses because they take place over short intervals with sparse cadence, typically yielding only $\sim$16 observations over a $\sim$36-hour interval~\citep{wrig10}.  Nevertheless, most asteroids experience a few rotations in 36 hr, such that WISE data can in principle be used to estimate the spin periods of thousands of asteroids \citep{cutr19}.

\citet{dure15} combined sparse photometry, including WISE data, to derive the spin period of
one asteroid.  \citet{hanu15} combined WISE thermal infrared data and other data to obtain shape models and spin periods of
six asteroids.  Their work has since been expanded to derive 1451 spin periods for asteroids observed by WISE \citep{dure18}.  Here, we use the well-curated data set from the fully cryogenic phase of the WISE mission \citep{myhr22,gold4} to estimate spin periods for hundreds of asteroids.

The Lightcurve Database (LCDB) \citep{lcdb} is a compilation of most known lightcurve measurements from various sources.  Each lightcurve is assigned a quality code
between 0 (incorrect) and 3 (best) by the database curators to convey the confidence level in the uniqueness and accuracy of the rotational period estimate.
We used the LCDB to evaluate the reliability of our solutions and to train a machine learning reliability classifier.

\section{Methods}
\label{sec-methods}

\subsection{Overview}
\label{sec-overview}

Our methods closely follow those of \citet{wasz15}, who used sparse photometry from the Palomar Transient Factory (PTF) to determine $\sim$9,000 reliable asteroid spin periods. Their method is conceptually straightforward. For each trial period, one fits a Fourier series model \citep[][Equation 1]{harr89} to the observed flux values and computes the sum of squares of the flux residuals. The Fourier series is truncated after the second harmonic, a simplification that rests on the assumption that the object is approximately ellipsoidal in shape.  It is also consistent with the fact that the second harmonic dominates asteroid lightcurves with amplitudes greater than 0.4 magnitudes \citep[][]{harr14}. Although this model is insufficient to capture the full details of the lightcurve, it is adequate to recover the spin period in most instances, as can be verified by comparing the solutions to high-quality (quality code 3- or above) solutions published in the lightcurve database (LCDB) \citep{lcdb}.  A similar method was also used by \citet{chan17} to determine 2780 reliable asteroid rotation periods from the PTF.

\citet{wasz15}’s method is directly applicable to WISE photometry, which typically contains
at least $\sim$16 observations of each asteroid
with a nominal 1.59 hr cadence 
over a $\sim$36 hr period. 
Although this observational
mode prevents the determination of spin periods for fast ($P<$ 3.17 hr) and
slow ($P>$ 72 hr) rotators, 
most asteroids have spin periods that are amenable to characterization with this technique.
Based on reliable LCDB statistics (quality code 3- or higher) and the range of diameters (0.28--72.2 km) in the \citet{myhr22} sample, we estimate that fewer than 17\% of asteroids in our sample have a spin period smaller than 3.17 hr.

\citet{wasz15} were able to fit for a photometric phase function because their observations were obtained over a wide range of phase angles.  In contrast, WISE observations typically span a narrow range of phase angles and we did not attempt to evaluate the phase function.  Phase angles remained nearly constant during the short observation intervals, and phase angle effects were absorbed by the zeroth-order coefficient of the Fourier series, i.e., mean magnitude.

We compared our method to results obtained with the more traditional Lomb-Scargle periodogram \citep{press92} (Sections~\ref{methods-ls} and \ref{res-ls}).

\subsection{Data Set}
\label{sec-data} 

\subsubsection{Initial Data Set}
\label{sec-data-gold4}

We used the carefully curated data set of \citet{myhr22}, who eliminated measurements with artifacts, low signal-to-noise ratio (S/N), poor photometric quality, saturation, questionable PSF fits, background confusion, problematic near-conjunction conditions, or large discrepancies between ephemeris predictions and reported position.  Their data set provides high-quality flux measurements in all four infrared bands for 4420 asteroids.  For a small fraction (6\%, 265 asteroids),
WISE observations were obtained in distinct (almost always two) epoch clusters separated by more than 30 days.
We determined independent solutions for each cluster of observations because they were obtained at different phase angles.
Although our analysis is done on individual clusters, we may refer to the clusters as asteroids for ease of presentation.

\subsubsection{Assignment of Flux Uncertainties}
\label{sec-sigma}

Both \citet{hanu15} and \citet{myhr18empirical} have shown that uncertainties reported by the WISE pipeline underestimate actual flux uncertainties.  \citet{hanu15} used $\sim$400 pairs of asteroid detections observed in quick succession ($\sim$11 s) to quantify actual flux uncertainties in W3 and W4.  They found that uncertainties reported in the WISE database underestimate actual uncertainties by factors 1.4 and 1.3 in W3 and W4, respectively.  \citet{myhr18empirical} expanded this analysis to bands W1--W4 and included a much larger number of pairs (7834, 11202, 125318, and 59049 in  W1--W4, respectively).  Here, we further expanded the number of pairs and fit Gaussians to the distributions of Z values $(Z=(f_1 - f_2)/\sqrt{\sigma_1^2+\sigma_2^2})$ \citep[][Equation 3]{myhr18empirical} after removing $\sim$1\% of pairs with $|Z|>5$.  The elimination is required because the tails of the distributions are non-Gaussian even though the cores of the distributions are well approximated by Gaussians.  We list the correction factors in all four bands for completeness (Table \ref{tab-factors}).  We assigned uncertainties to the flux measurements by multiplying the uncertainties reported in the WISE database with the relevant correction factor.

\begin{table}[h]
  \begin{center}
    \begin{tabular}{lrrr}

    Band & Number of pairs & Correction factor & Median $\sigma$ (corrected) \\
    \hline
 W1 & 29936  & 1.224 & 0.195 \\
 W2 & 34462  & 1.120 & 0.132 \\
 W3 & 170232 & 1.479 & 0.035 \\
 W4 & 120225 & 1.218 & 0.058 \\
  \end{tabular}
    \caption{Correction factors that are required to convert flux uncertainties reported in the WISE pipeline to actual uncertainties.  The second column shows the number of remaining pairs after elimination of $\sim$1\% of pairs with uncharacteristically large flux differences.
      The last column shows the median flux uncertainties in magnitude units after correction.
      }
    \label{tab-factors}
\end{center}  
  \end{table}

\subsubsection{Band Selection}
\label{sec-band-selection}

\begin{figure}[hbt]
\begin{center}
  \begin{tabular}{cc}
    \includegraphics[width=3in]{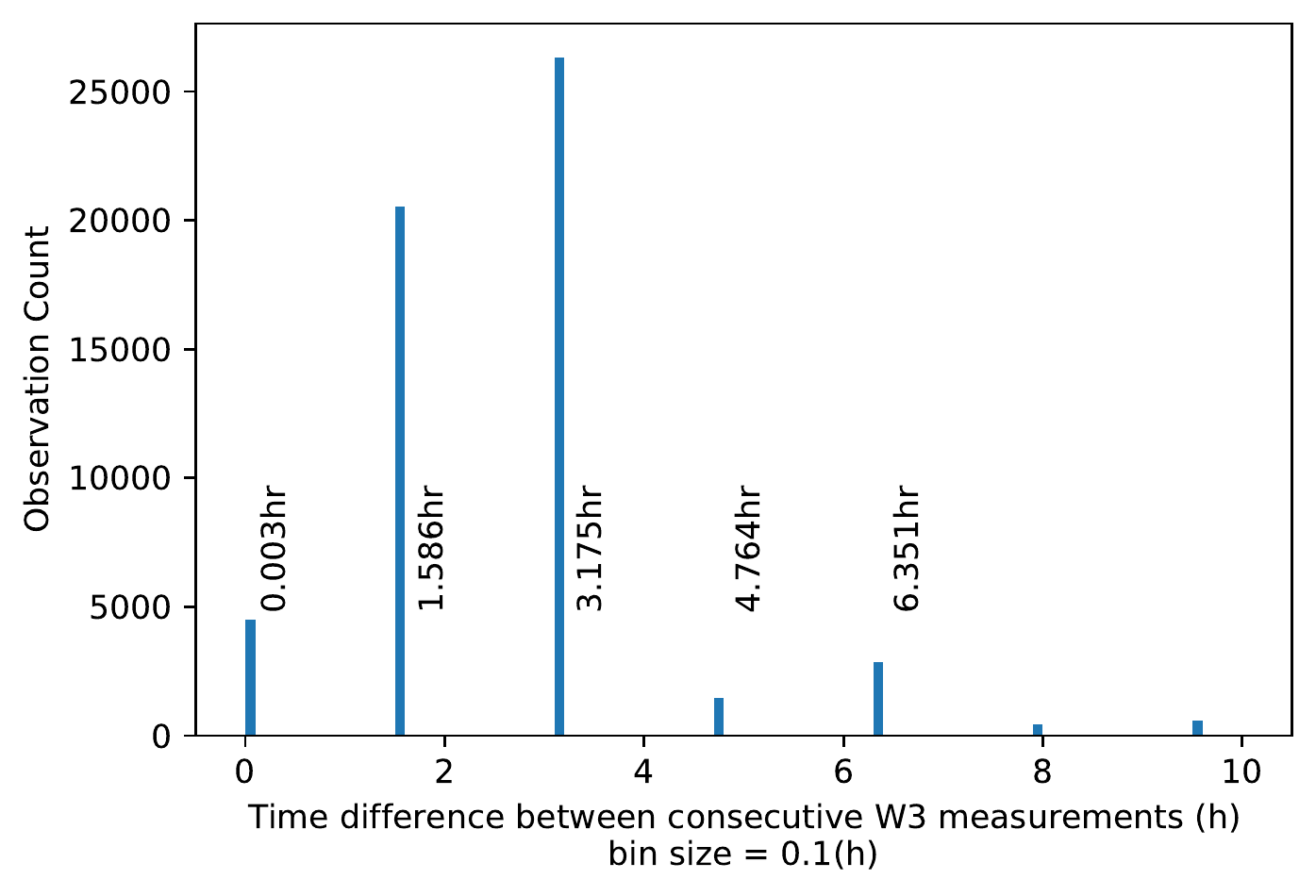} & \includegraphics[width=3in]{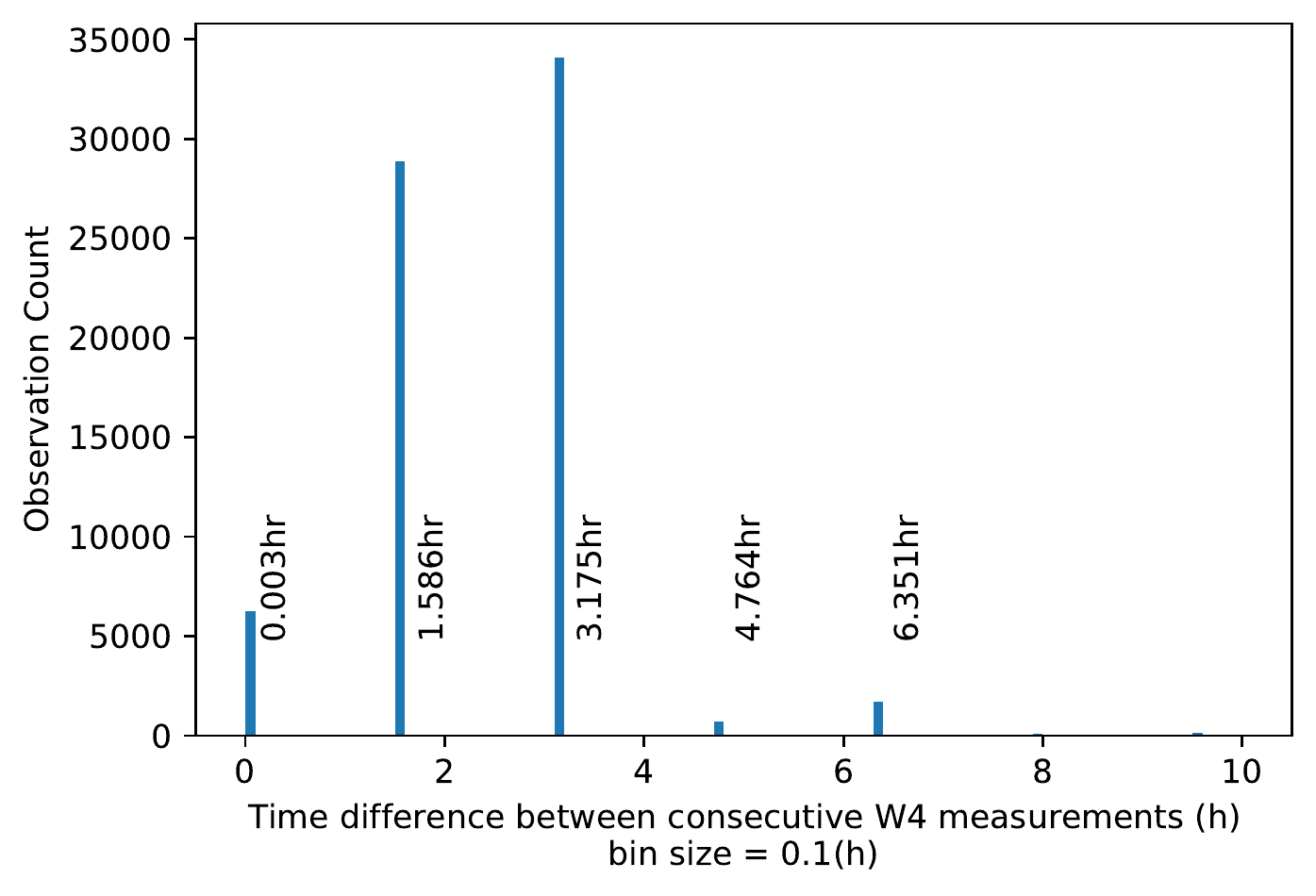}
  \end{tabular}
  \caption{Histograms of W3 (left) and W4 (right) sampling intervals in the \citet{myhr22} data set.}
\label{fig-w3-cadence}
\end{center}
\end{figure}

We focus on observations in W4 for two reasons.  First, thermal emission from asteroids is stronger, S/N is higher, and the number of observations in the curated data set is higher in W3 and W4 than in W1 and W2.  
Second, the observation cadence of the original and curated WISE data 
is generally non-uniform, with time intervals as small as 11~s, a most common time interval of 1.59 hr, a frequent time interval of 2 $\times$ 1.59 hr = 3.17 hr, and occasional time intervals at higher multiples of 1.59 hr.
W4 observations provide the best observational cadence, whereas W3 observations have a
higher fraction
of asteroids with a longer cadence (Figure \ref{fig-w3-cadence}), which increases the susceptibility to aliasing difficulties (Section {\ref{sec-alias}}).

We are not concerned with thermal lags or differences in lightcurve amplitudes compared to optical lightcurves, as they are inconsequential to the determination of spin periods. In addition, \citet{dure18} demonstrated that thermal lightcurves in W3 and W4 are qualitatively consistent with optical lightcurves.  

While multiband lightcurve fits could in principle be envisioned, their utility in the context of spin period determinations with WISE data is limited because the exposures in the four bands are simultaneous, such that they sample roughly the same rotational phase.  The determination of phase lags among observations in the four WISE bands is potentially informative but beyond the scope of the current work.

\subsubsection{Data Selection Filters}
\label{sec-data-prefilters}

We applied several preprocessing filters in order to identify lightcurves most suitable for rotational period determination. Each observation cluster must pass all of the following filters to qualify for analysis, otherwise it is discarded.
First, we focus on
lightcurves that have at least 12 data points in W4.
Second, we eliminated
lightcurves where the peak-to-peak variation in flux magnitude W4$_{{\rm red}} < 0.3$ mag,
where we used magnitudes reduced to the values that would be observed at an asteroid-sun distance $r_{\rm as}$ = 1 au and asteroid-observer distance $r_{\rm ao}$ = 1 au.
This filter eliminates low-amplitude lightcurves, which may be ambiguous with respect to spin period determination \citep{harr14}.
This filter effectively eliminates very slow rotators, which are not considered in our analysis anyway (Section \ref{sec-fit}).
Third, we eliminated any
lightcurve that does not have an adequate observation cadence.  Specifically, we required at least one time interval between consecutive observations to fall in the range 1.55--1.58 h, which corresponds to the WISE spacecraft orbital period \citep{wrig10}.  Observation clusters that have a longer (3.10--3.16 h)
minimum sampling interval are more susceptible to severe aliasing (Section \ref{sec-alias}) and are discarded.

The filtered data set contains a total of 3061 asteroids, with 3225 observation epochs and 57,532 W4 photometric measurements.  In this data set, 164 (5.3\%) out of the 3061 asteroids contain two observation clusters, and none have more than 2 observation clusters.

On average, an observation cluster contains 16 data points and spans 36 hr.

\begin{table}[h]
  \begin{center}
    \begin{tabular}{lr}
    W4 flux observations & 57,532 \\ %
     Median flux uncertainty & 0.067  \\ %
     Median number of data points per cluster & 16 \\
     Median observation span of clusters & 1.522 days\\

  \end{tabular}
      \caption{Characteristics of filtered data set}
  \label{tab-preproc-stats}
\end{center}  
  \end{table}

\subsection{Fitting Procedure}
\label{sec-fit}

An effective method for identifying periodicities in sparse data is a ``period scan'', where a model lightcurve is fitted to the data for a range of trial spin periods and a measure of the misfit (i.e., fit dispersion) is evaluated at each trial period \citep{harr12}.  The misfit metric is the usual sum of squares of residuals:
\begin{equation}
\chi^2 = \sum_{i=1}^{N} \frac{(O_i - C_i)^2}{\sigma_i^2},
\end{equation}
where $O_i$ is the $i$-th observation, $C_i$ is the $i$-th computed (modeled) value, $\sigma_i$ is the uncertainty associated with the $i$-th observation, and the index $i$ ranges from 1 to $N$, the total number of observations.  The observations are the reduced flux magnitudes (Section \ref{sec-data-prefilters}), the uncertainties are the magnitude uncertainties from the WISE database multiplied by the relevant correction factors (Section \ref{sec-sigma}), and the computed values are obtained by fitting a second-order Fourier model to the data similar to \citet{wasz15}'s formulation.  Specifically, 
\begin{equation}
  C_i = A_{0,0}
      + A_{1,1} \sin{\left(\frac{2\pi}{P} t_i\right)}
      + A_{2,1} \cos{\left(\frac{2\pi}{P} t_i\right)}
      + A_{1,2} \sin{\left(\frac{4\pi}{P} t_i\right)}
      + A_{2,2} \cos{\left(\frac{4\pi}{P} t_i\right)},
      \label{eq-fourier}
\end{equation}
where $t_i$ is the light-time corrected epoch of the $i$-th observation, $P$ is the trial spin period, and the $A_{i,j}$ are the five adjustable Fourier coefficients.

We considered a range of evenly spaced rotational frequencies between $f$ = 0 and 7.5662 rotations per day, where $f_{\rm max}$~=~7.5662 rot/day ($P \simeq 3.17$ h) represents the fastest rotation that can be
Nyquist sampled with the WISE observational cadence (Section \ref{sec-alias}).
However, we showed that it is possible to recover the periods of fast rotators by taking advantage of the mirroring properties of aliased signals.
We did not attempt to recover rotational frequencies that exceed 11 rotations per day, which correspond approximately to the rotational frequency at which centrifugal acceleration at the equator exceeds the acceleration due to self-gravity for typical asteroid densities, the so-called spin barrier at $P\simeq 2.2$ hr \citep{prav02}.  The overwhelming majority of asteroids detected by WISE and included in our data set are large and experience fewer than 11 rotations per day.

The least-squares minimizer is an ordinary linear least-squares solver that follows \citet[][tar.gz file]{wasz15}, except that we do not fit for phase curve parameters.
We also computed the reduced chi-squared metric $\chi^2_\nu = \chi^2 / (N-5)$, where the number of degrees of freedom $\nu = N-5$ represents the number of data points minus the number of free model parameters. %

The discrete Fourier transform of a time series with duration $T$ yields a frequency resolution of $1/T$.  In this work, we chose to increase the frequency resolution by an oversampling factor of 10 in order to better resolve peaks in the %
period scan.  The number of trial frequencies was therefore set to $10 \times T \times f_{\rm max}$.

We used an iterative procedure similar to \citet{wasz15}
where an increasingly large ``cosmic error'' is added to the observation uncertainties.
The cosmic error is initialized as 0.002 mag in the first iteration and is multiplied by 1.5 in each subsequent iteration. %
The purpose of the cosmic error is to inflate the measurement uncertainties in order to reflect the model's inability to accurately represent asteroid lightcurves with a Fourier series truncated at the second harmonic.  The cosmic error does not
affect the periodicities identified in the lightcurve, but it does affect the confidence intervals assigned to the period estimates.

We followed \citet{wasz15} and \citet{harr14} in preferring double-peaked folded lightcurves. 
To identify the number of peaks in the lightcurve, we generated a synthetic folded lightcurve with the fitted Fourier coefficients and candidate period, sampled it with 10,000 points, and analyzed the samples with Matlab's built-in function {\tt find\_peaks}\footnote{https://www.mathworks.com/help/signal/ref/findpeaks.html}.
For each double-peaked solution, we computed the heights of each peak relative to the lightcurve's global minimum.

There are two possible paths to convergence.  At the end of each iteration, the solution with the lowest $\chi^2_\nu$ is selected if and only if it satisfies three conditions: (1) the folded lightcurve is double-peaked;
(2) the height of the highest peak is at least twice that of the lowest peak; and (3) $\chi^2_\nu < 3$.  If conditions (1) or (2) are not satisfied, the solution at half-frequency is considered and is adopted if it satisfies the same three conditions.  Otherwise, the cosmic error is increased
and the next iteration begins.
If the cosmic error reaches 0.1 mag, the fit is deemed unsuccessful.

\subsection{Aliasing}
\label{sec-alias}

The cadence of observations determines the sampling intervals between consecutive photometric measurements.
The Nyquist sampling criterion requires that at least two samples of a periodic signal be obtained per cycle in order to identify the periodicity unambiguously.
Apart from the fortuitous double detections obtained $\sim$11 s apart (Section \ref{sec-sigma}), the smallest sampling interval in WISE data is approximately 1.59 hr, which is dictated by WISE's $\sim$15 daily orbital revolutions \citep{wrig10}.
Asteroids sampled with the 1.59 hr cadence and rotation periods larger than 3.17 hr ($f <$7.5662 rot/day) are usually Nyquist sampled, i.e., suffer no aliasing.

Assuming uniform sampling, the signatures of asteroids with rotation periods between 1.59 hr and 3.17 hr appear aliased in the period scan in a predictable manner, specifically:
\begin{equation}
  f_{\rm alias} = 2 f_{\rm Nyq} - f_{\rm spin},
  \label{eq-nyquist}
\end{equation}
where $f_{\rm Nyq}$ = 7.5662 rot/day is the critical Nyquist frequency or folding frequency, $f_{\rm spin}$ is the underlying true rotational frequency, and $f_{\rm alias}$ is the aliased frequency.
For example, an asteroid rotating at the spin barrier of 2.2 h (10.9 rot/day) exhibits signatures at 2.2 h (10.9 rot/day) and 5.66 h (4.24 rot/day). 
Absent additional information, a period scan may yield inconclusive results with respect to these two solutions. 
However, folding about the 7.5662 rot/day axis remains limited because only about 17\% of asteroids in our sample experience rotation rates that exceed 7.5662 rot/day (Section \ref{sec-overview}).
The overwhelming majority of asteroids detected by WISE and included in our data set are large and experience fewer than 11 rotations per day.

It is frequent
for the interval between consecutive W4 measurements to be 3.17 hr instead of the nominal cadence of 1.59 hr, e.g., when poor-quality flux measurements are eliminated.  As a result, we also expected and observed
(Section~\ref{sec-results})
folding of the periodogram about  $f_{\rm Nyq'}$ = 3.7831 rot/day.
The folded frequency happens to be correct in a substantial fraction of cases, which we used to our advantage as it is calculable and therefore recoverable with no loss of precision.
Additional aliasing considerations are described in Appendix~\ref{app-aliasing}.

\subsection{S/N Calculations}
\label{sec-snr}
A period scan may return multiple peaks with low misfit values.  S/N metrics are useful in determining whether a peak is likely to represent a genuine rotational signature as opposed to a noise artifact.  We adopted two S/N metrics.  One metric follows \citet{wasz15} and quantifies the height of the peak with respect to the median misfit in terms of an estimation to the standard deviations of the misfit variations:
\begin{equation}
  S/N_{\rm W} = \frac{|\chi^2_{\rm min} - \chi^2_{\rm median}|}{(\chi^2_{84\%} - \chi^2_{16\%})/2}, 
  \label{eq-snrw}
\end{equation}
where the denominator includes percentiles of the misfit distribution corresponding to $\pm$1 standard deviations from the median.
The second metric follows \citet{harr12} and associates the misfit outside of minima, which we approximated by $\chi^2_{95\%}$, to the quadratic sum of the amplitude of lightcurve variation ($a$) and the noise in the data ($n$).  It also associates the minimum misfit to the square of the single-point data scatter after removal of the signal ($n^2$), and assigns an overall noise level to the solution equal to $n/\sqrt{\nu} = n/\sqrt{(N-5)}$.  We have
\begin{equation}
  (a^2 + n^2) = \chi^2_{95\%} / N,
\end{equation}
  \begin{equation}
    n^2 = \chi^2_{\rm min} / N,
\end{equation}
  \begin{equation}
    {\rm signal} = a =  \sqrt{(\chi^2_{95\%} - \chi^2_{\rm min}) / N}
  \end{equation}
  \begin{equation}
    {\rm noise} = n' = \sqrt{\chi^2_{\rm min} / N} / \sqrt{(N - 5)}
  \end{equation}
  \begin{equation}
    S/N_{\rm H} = a/n' = \sqrt{\frac{\chi^2_{95\%} - \chi^2_{\rm min}}{\chi^2_{\rm min} }} \sqrt{(N - 5)}
    \label{eq-snrh}
  \end{equation}

\subsection{Assignment of Spin Period Uncertainties}
\label{sec-uncertainties}

Once the best-fit peak was identified, we assigned a 1$\sigma$ uncertainty to the fitted period by computing the periods corresponding to a constant chi-square boundary \citep[][Section 15.6]{press92}.  Specifically, we computed the periods at which
$ \chi^2 = \chi^2_{\rm min} + \Delta\chi^2(68.3\%, \nu=5) = \chi^2_{\rm min} + 5.86$, where  $\chi^2_{\rm min}$ is the minimum misfit.

\subsection{Post-processing Filters}
\label{sec-post-fit-filters}
Because the duration and cadence of WISE observations are not optimal for the unambiguous determination of spin periods, it was important to
remove solutions that are
likely unreliable.
We applied the following filters:

(1) Reject slow rotators.  
We required observations over at least 180 degrees of rotational phase and rejected any solution with a best-fit period that is two or more times longer than the data span.
(2) Reject anomalously high amplitude lightcurves.
We eliminated solutions where the peak-to-peak amplitude of the fitted solution was three or more times larger than the peak-to-peak amplitude of the observations.
The peak-to-peak amplitude of the lightcurve was determined numerically while evaluating the lightcurve with the best-fit Fourier coefficients.

(3) Reject low S/N solutions.  
  Low S/N solutions are likely spurious and were eliminated.  In practice, we found that the S/N formulation of \citet{harr12} was more effective than that of \citet{wasz15}, perhaps due in part to the relatively small number of (noisy) observations. 
  Solutions were rejected when $(S/N)_{\rm H} < 5$.
  The solutions that passed all of the above filters are reported below. 
  We have evidence that most of these solutions are accurate
  (Section~\ref{sec-results}),
  but we also expect a fraction of aliased or incorrect solutions in this set.

\subsection{Machine Learning Reliability Classifier} 
\label{sec-ml}

\citet{wasz15} pioneered the usage of a machine learning (ML) classifier to improve the reliability of asteroid lightcurve fits.
They applied a random forest (RF) algorithm, which is 
a supervised machine learning algorithm that utilizes an ensemble of weak decision tree predictors to increase prediction power. 
The hypothesis underpinning a machine learning classifier is that certain appropriately chosen features associated with a lightcurve solution jointly carry non-trivial information regarding the reliability of the solution.  Given a labeled training set that includes both the values of the features and a reliability indicator, an ML algorithm can be trained to detect relations within the feature space and predict a reliability indicator for lightcurve solutions that do not have a reliably known period (i.e., solutions that are not in the training set).
A Random Forest classifier makes predictions via a majority voting process by its ensemble of decision trees. For each sample, the classifier generates a probability derived from the voting process, then makes a binary prediction (i.e., reliable or unreliable) on the basis of a user-defined probability threshold. 
\citet{wasz15}'s classifier was trained with about 1000 lightcurves with known reference periods and improved the overall
success rate 
from $\sim$66\% to $\sim$80\% for 19,000 lightcurves. %

Because our work also involved the analysis of thousands of sparse lightcurve, we initially applied an RF algorithm in an attempt to identify the most reliable lightcurve solutions.  The RF classifier was able to provide a modest improvement to the success rate of our primary solutions (from 55\% to 70\%), but it also marked correct solutions as incorrect.  Because the recovery of spin periods among our three solutions was so high (88\%) and the performance of the RF classifier was limited, we ultimately decided against providing a potentially flawed reliability indicator.

\subsection{Lomb-Scargle Periodogram}
\label{methods-ls}

The Lomb-Scargle (LS) periodogram \citep{Lomb76,Scar82} is a standard algorithm that enables the analysis of periodicities in unevenly sampled time series, 
such as asteroid lightcurves. 
We explored the performance of the LS periodogram
as an alternative to our default pipeline.  Our LS pipeline follows the implementation in standard libraries and does not include iterative adjustment of uncertainties, requirement for double-peaked solutions, and post-fit filters implemented in our default pipeline.
We used the generalized LS algorithm of \citet{zech09} as implemented in the astropy package\footnote{https://docs.astropy.org/en/stable/timeseries/lombscargle.html}.  This implementation takes observational uncertainties into account and enables the estimation of a floating mean.
We deployed both first-order and second-order LS periodograms.

In the first-order LS implementation, we followed \citet{McNeill19}
and set the estimated rotational period at twice the best-fit LS period to conform to the double-peak nature of asteroid lightcurves.
We admitted only solutions with a false-alarm probability (FAP) less than 10\%. %
The second-order LS implementation is similar to the periodogram calculation used in our default pipeline but does not include any post-fit filter.  

\section{Results}
\label{sec-results}

\subsection{Default pipeline}

We present the period solutions that successfully converged in the Fourier fitting algorithm and passed the post-fit filters (Section~\ref{sec-post-fit-filters}). 
The period solutions of 2008 ($\sim$62\%) out of the initial 3225 lightcurves fulfilled both inclusion criteria.

To test the reliability of our results, we compared our spin period estimates to high-quality (quality code 3- or higher) rotational periods published in the LCDB.
We quantified the fraction of solutions that were within 5\% of the LCDB solution, which are deemed to be accurate solutions.
At the time of writing, there were 752 solutions (representing 702 unique asteroids) among our 2008
solutions with a suitable LCDB estimate.
We refer to this set of solutions as the `LCDB reference group'.

\begin{figure}[hbt]
 \begin{center}
     \includegraphics[width=4in]{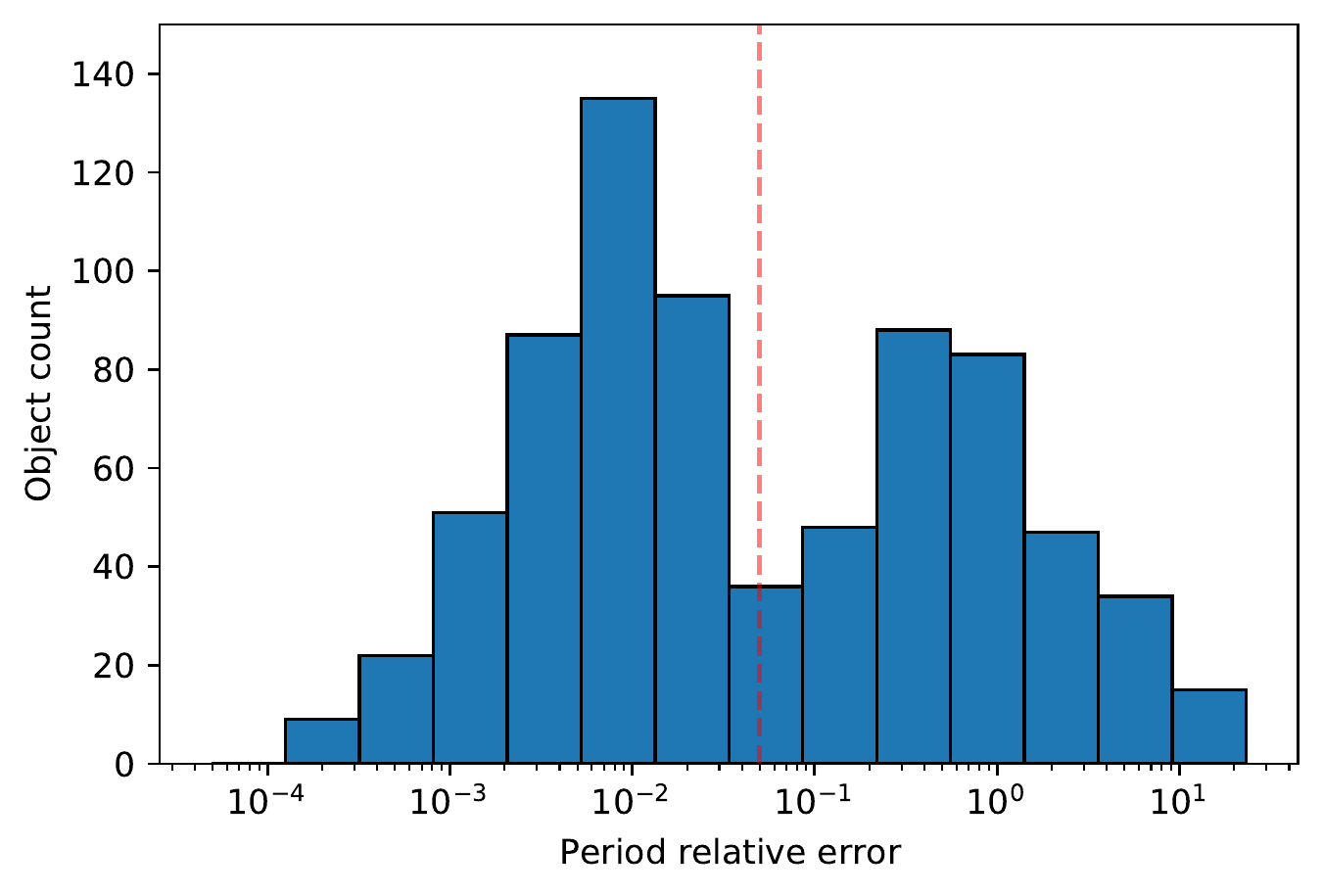}
     \caption{Histogram of fractional accuracy in spin period obtained by comparing our period solutions to the corresponding LCDB reference periods.  The left and right clusters correspond to the accurate and inaccurate period estimates, respectively. The red line denotes the 5\% accuracy threshold.}
\label{fig-accuracy-histogram}
\end{center}
\end{figure}

In the LCDB reference group, the fitted period was found to be accurate (within 5\%) in
55\% of the cases. Notably, the relative errors of the fitted periods exhibit a bimodal distribution (Figure \ref{fig-accuracy-histogram}).
The two
modes bifurcate at a fractional error of approximately 5\%,
which provides a posteriori justification for selecting a 5\% threshold for accuracy.

\begin{figure}[p]
 \begin{center}
     \includegraphics[width=7in]{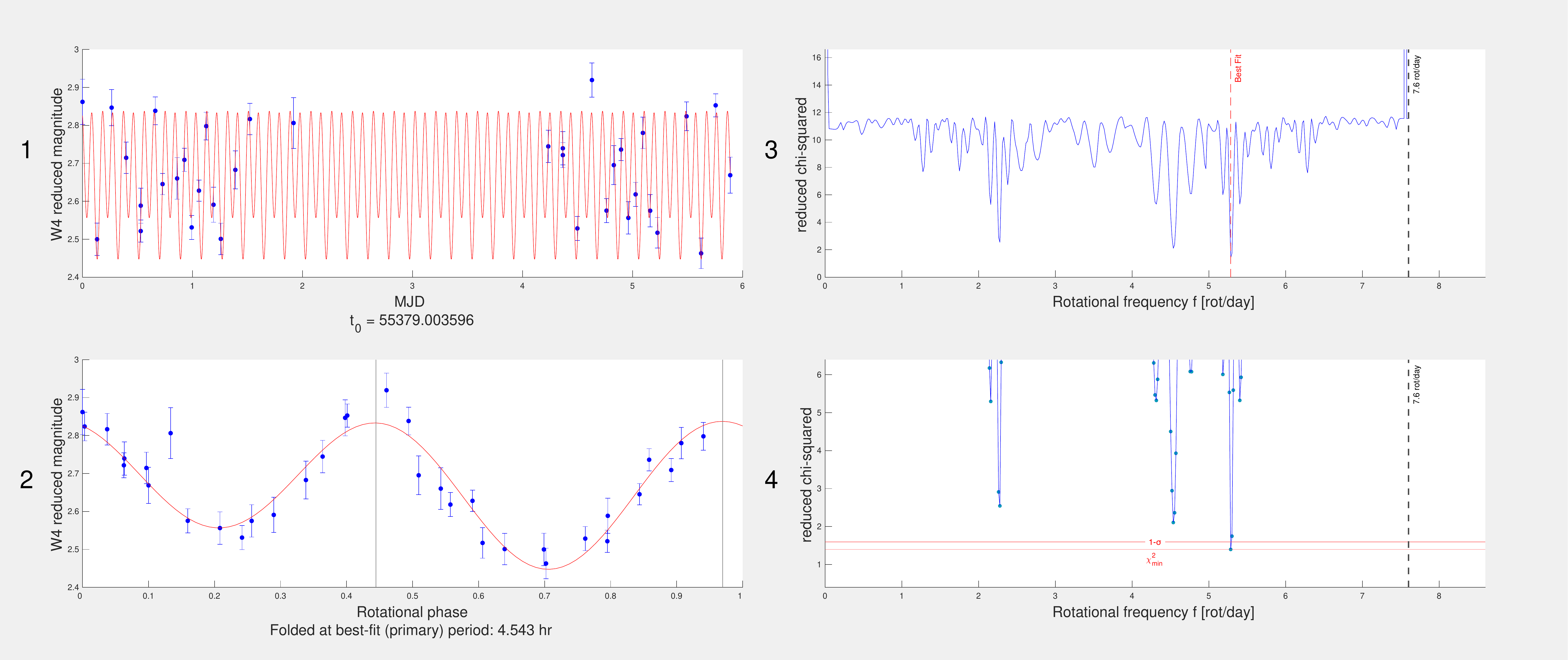}
     \caption{
       Spin period solution for asteroid 296 obtained from 35 W4 observations spanning 6 days.  
       The period solution at 4.543 $\pm$ 0.014 hr is in good agreement with the LCDB value of 4.5385 hr.
       The data exhibit short (1.59 h) sampling intervals over a long observation span, and the periodogram is free of aliases.
     }
\label{fig-suc-mode-1}
\end{center}
\end{figure}

\begin{figure}[p]
 \begin{center}
     \includegraphics[width=7in]{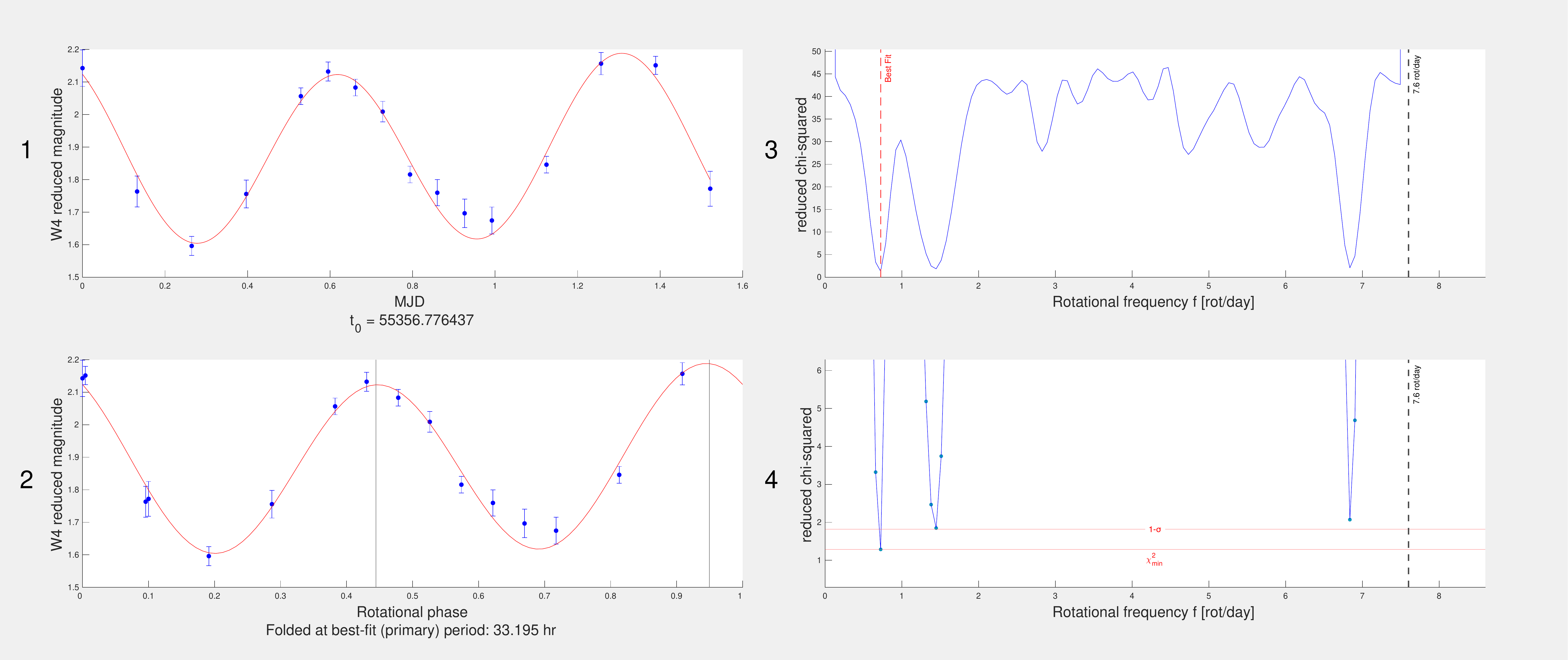}
     \caption{
       Spin period solution for asteroid 2715 obtained from 16 W4 observations spanning 1.5 days.
       The period solution at 33.195 $\pm$ 2.19 hr is in good agreement with the LCDB value of 33.62 hr.
       The correct solution was identified despite a data observation span that is only slightly longer than the spin period.
     }
\label{fig-suc-mode-3}
\end{center}
\end{figure}

\begin{figure}[p]
 \begin{center}
     \includegraphics[width=7in]{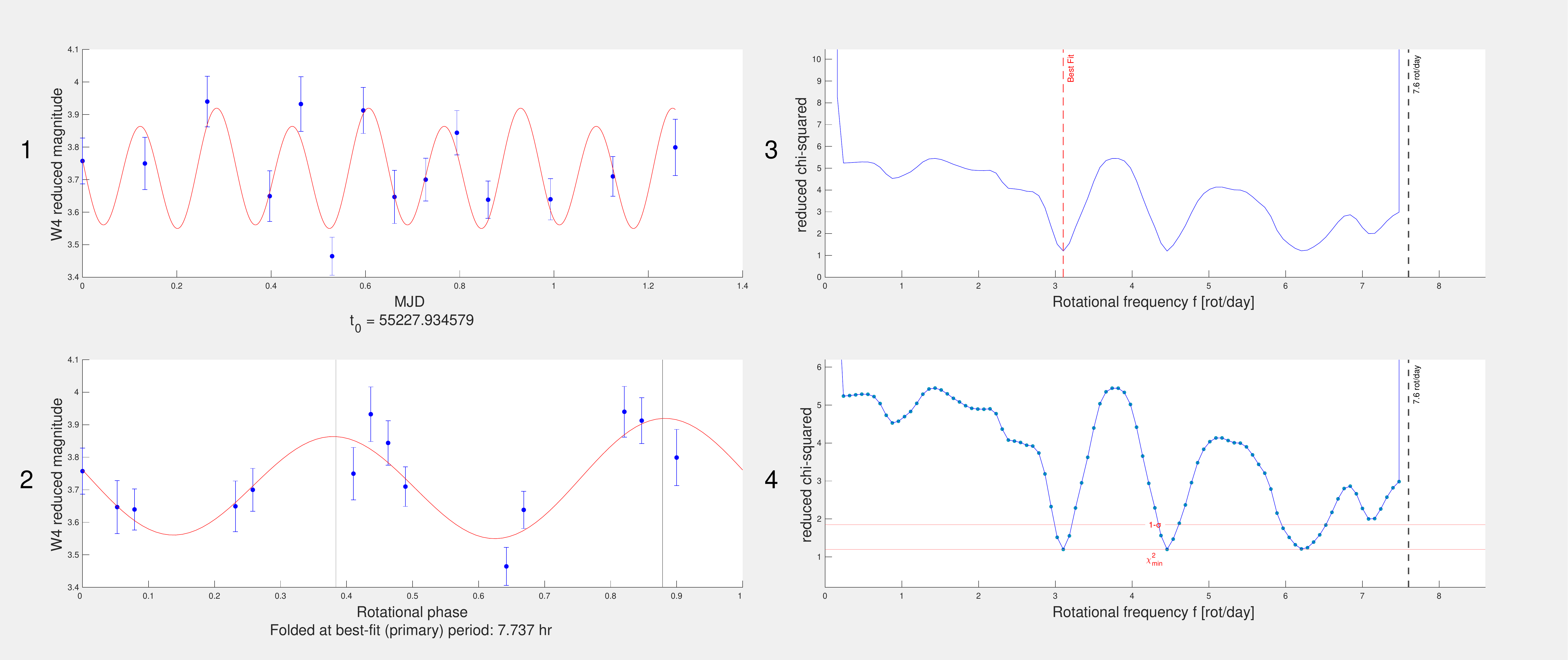}
     \caption{
       Spin period solution for asteroid 2812 obtained from 14 W4 observations spanning 1.2 days.
       The period solution at 7.737 $\pm$ 0.30 hr is in good agreement with the LCDB value of 7.7 hr.
       The correct solution was identified despite a relatively low S/N and the presence of competing solutions with similar but larger $\chi^2$ values.
            }
\label{fig-suc-mode-2}
\end{center}
\end{figure}

Our spin period solutions are listed in Table~\ref{tab-results}.  Figure~\ref{fig-suc-mode-1} illustrates an example of a favorable situation with a short (1.59 hr) cadence and relative long (6 days) duration, which yields a solution with high S/N and no aliasing.  We found that it is possible to successfully recover the spin period even when the observation span is comparable to the spin period (Figure~\ref{fig-suc-mode-3}).  Our analysis also identified correct spin periods in situations where the minimum $\chi^2$ value is not markedly different from other competing solutions  (Figure~\ref{fig-suc-mode-2}).  When multiple periodogram peaks have comparable $\chi^2$ values, the potential for an incorrect solution exists.

\begin{figure}[p]
 \begin{center}
     \includegraphics[width=3in]{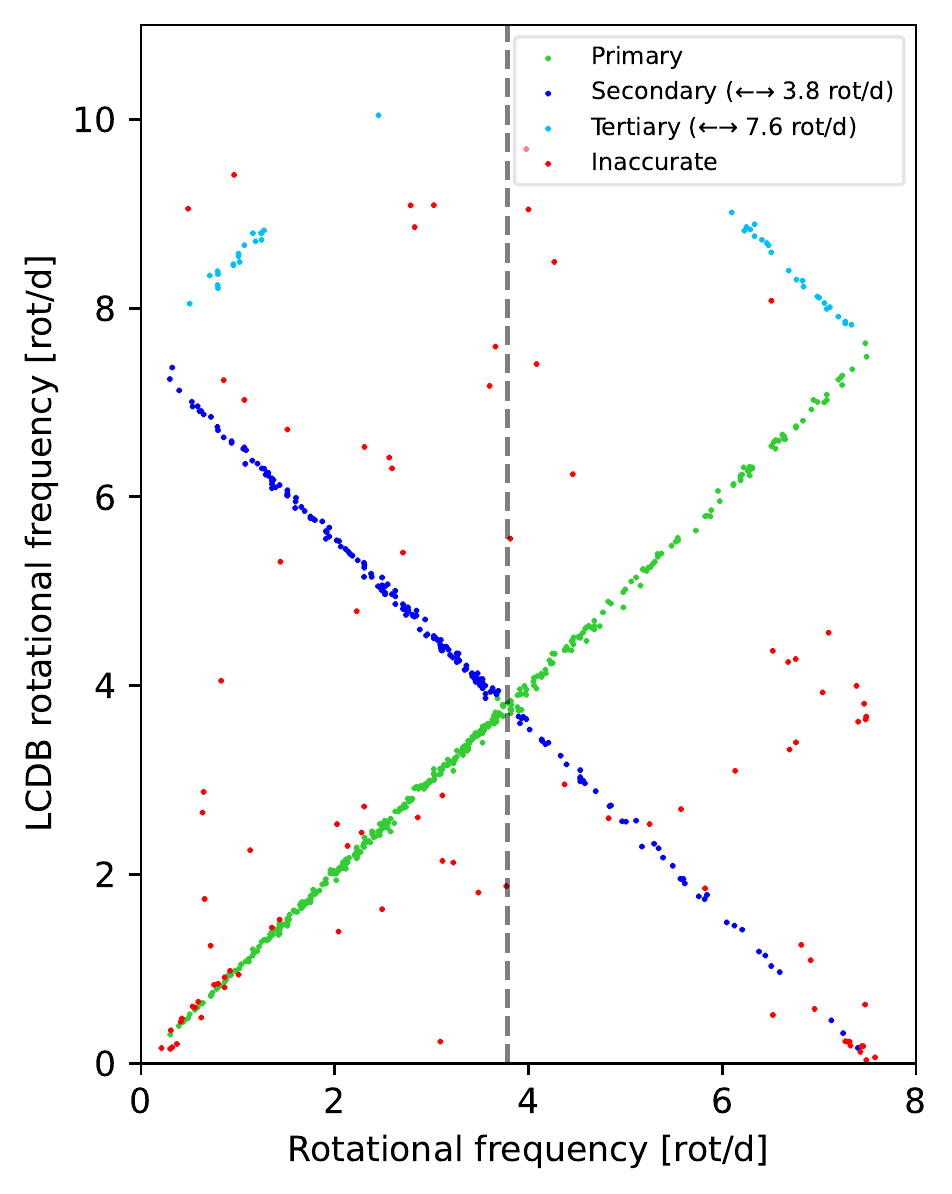}
     \caption{
              Best-fit rotational frequency (this work) vs.\ LCDB
              rotational frequency for the LCDB reference group.  The
              dashed grey line is at f = 3.7831 rot/day, which is the
              expected folding frequency for a 3.17 hr cadence and the
              predominant aliasing mode.  Most of the best-fit
              solutions (55\%) are accurate (green dots).  The
              solutions mirrored about 3.7831 rot/day (dark blue dots)
              are also accurate in 27\% of the cases, and the
              solutions mirrored about 7.5662 rot/day (light blue dots)
              are accurate in 6\% of the cases.  The inaccurate
              solutions (red dots) represent 12\% of the cases.
               The combination of best-fit and
              mirrored solutions yields an aggregate success rate of
       88\%.
}
\label{fig-pp-x-plot}
\end{center}
\end{figure}

\begin{figure}[p]
 \begin{center}
     \includegraphics[width=5.8in]{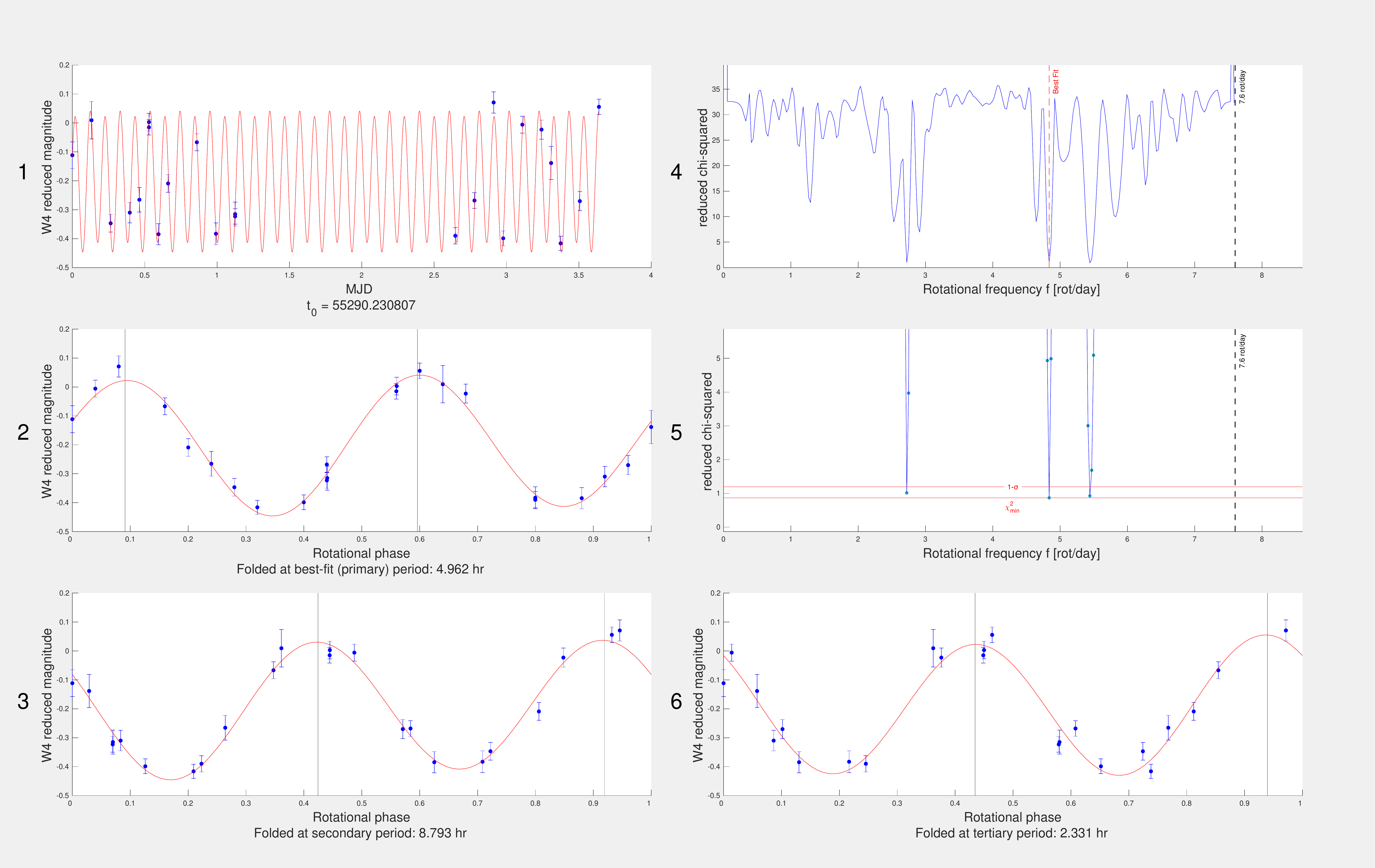}
     \caption{
       Spin period solution for asteroid 4715 obtained from 23 W4 observations spanning 3.6 days.
       The best-fit, primary period solution at 4.962 $\pm$ 0.01 hr (4.84 rot/day) is an alias (mirrored around f = 3.7831 rot/day) of the presumed correct LCDB value of 8.8129 hr (2.72 rot/day).  The secondary period solution at 8.7930 hr is accurate at the 0.2\% level.
          }
          \label{fig-mirror-mode-1}
\end{center}
\end{figure}

\begin{figure}[p]
 \begin{center}
     \includegraphics[width=5.8in]{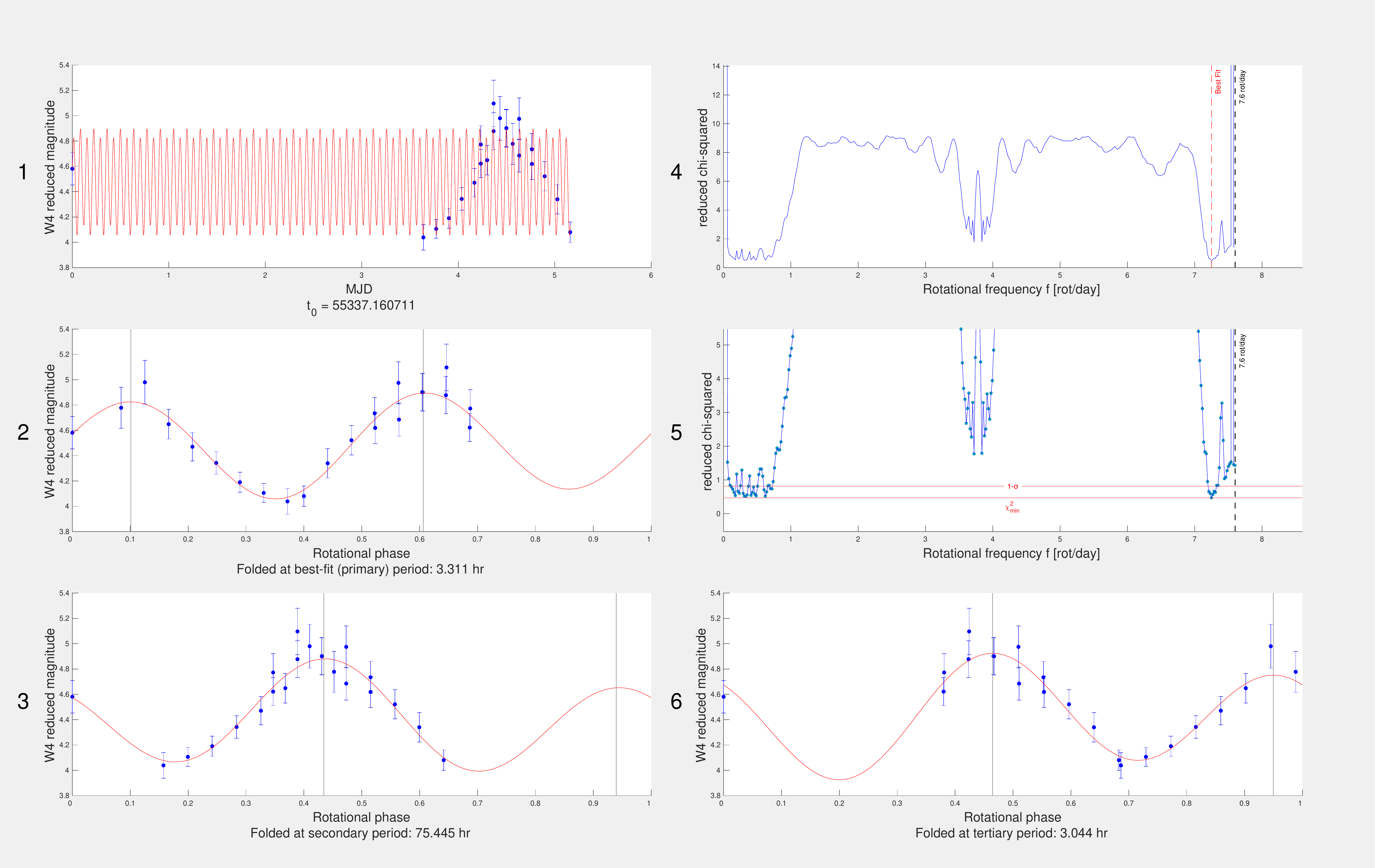}
     \caption{
       Spin period solution for asteroid 6485 obtained from 22 W4 observations spanning 5.2 days.
       The best-fit, primary period solution at 3.311 $\pm$ 0.035 hr (7.25 rot/day) is an alias (mirrored around f = 3.7831 rot/day) of the presumed correct LCDB value of 75.56 hr (0.318 rot/day).  The secondary period solution at 75.445 hr is accurate at the 0.2\% level.
}
\label{fig-mirror-mode-2}
\end{center}
\end{figure}

\begin{figure}[hbt]
 \begin{center}
     \includegraphics[width=7in]{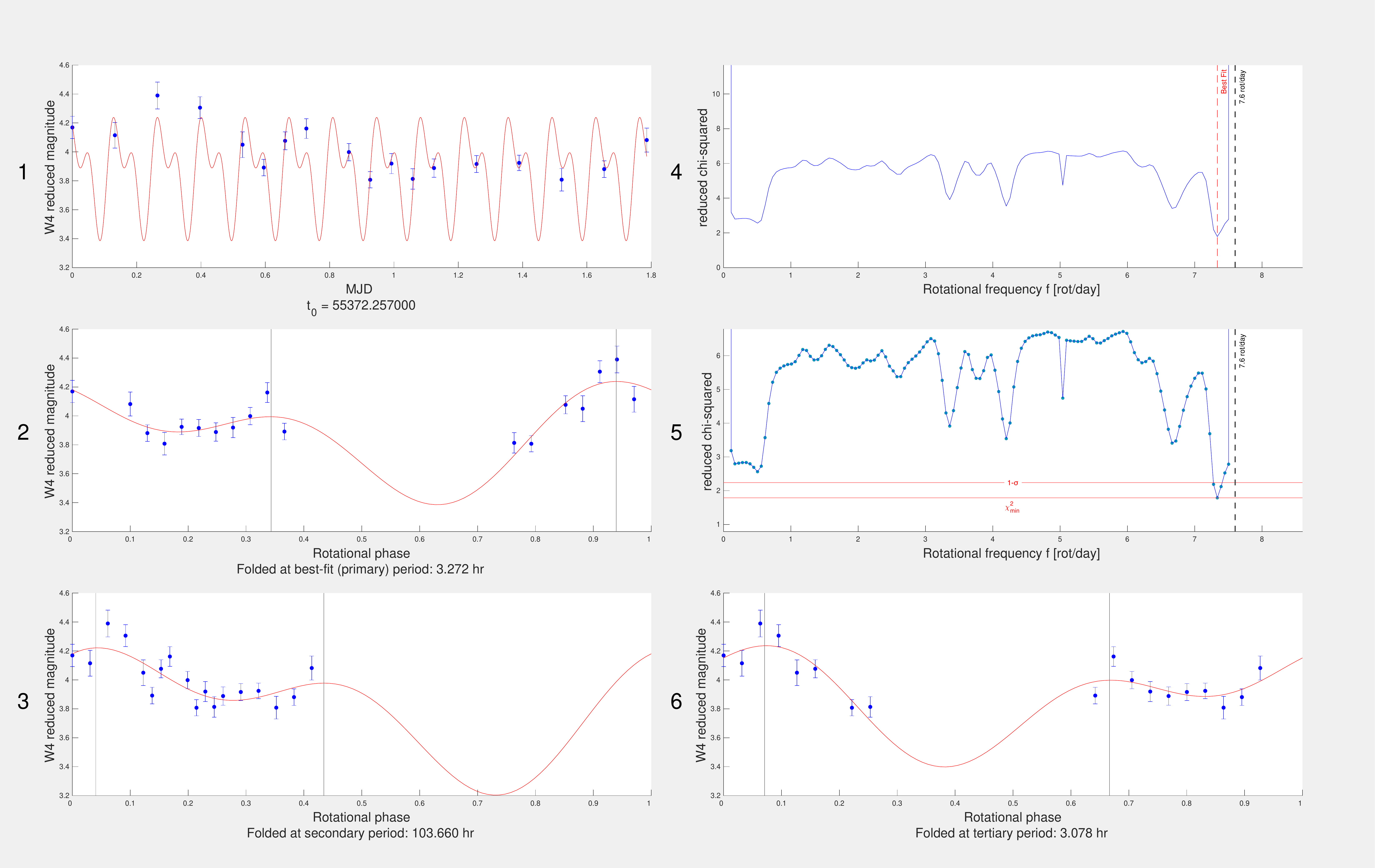}
     \caption{
            Spin period solution for asteroid 10721 obtained from 18 W4 observations spanning 1.8 days.
            The period solution at 3.272 $\pm$ 0.031 hr (7.33 rot/day) is an alias (mirrored around f = 7.5662 rot/day) of the presumed correct LCDB value of 3.0675~hr (7.82 rot/day).  The tertiary period solution at 3.078 hr is accurate at the 0.3\% level.
}
\label{fig-mirror-mode-3}
\end{center}
\end{figure}

We validated our
results by plotting our best-fit rotational frequencies against LCDB values for the LCDB reference group (Figure \ref{fig-pp-x-plot}).
In frequency space, which reveals folding behavior, the structure of the solutions is striking.  Most of the inaccurate solutions are in fact aliases of the
correct frequencies folded about the f = 3.7831 rot/day axis or, less frequently, the f~=~7.5662 rot/day axis.
Figures~\ref{fig-mirror-mode-1}, \ref{fig-mirror-mode-2}, and \ref{fig-mirror-mode-3} illustrate three examples.

\begin{figure}[hbt]
 \begin{center}
     \includegraphics[width=7in]{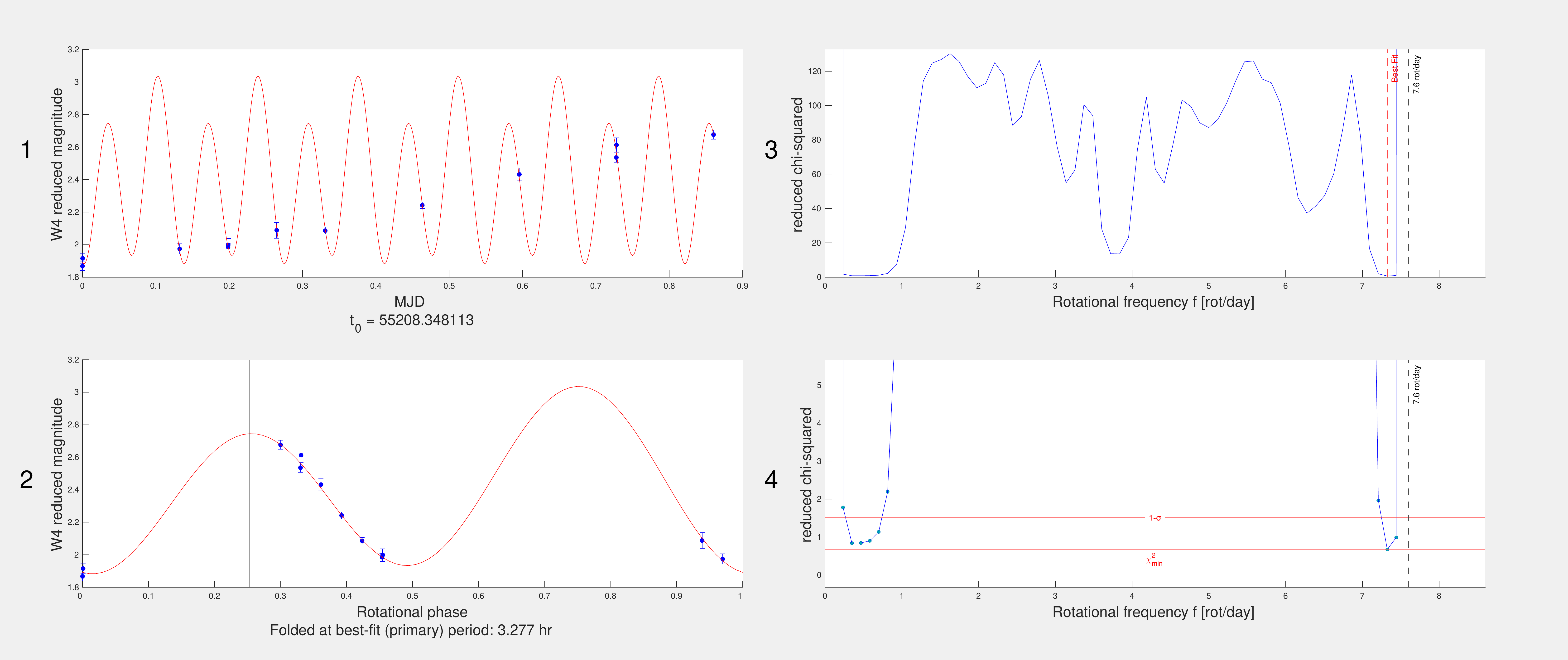}
     \caption{
       Incorrect spin period solution for asteroid 244 obtained from 12 W4 observations spanning 20.6~hr.  None of our solutions match the LCDB period (129.5 hr or 5.4 days).  WISE observations of this slow rotator cover only a small fraction of the full rotational phase, which prevents accurate recovery of the spin period.
}
\label{fig-failure}
\end{center}
\end{figure}

For this reason, we list both the best-fit frequency and its mirror values in Table~\ref{tab-results}.  One of these solutions is correct (within 5\%) in
88\% (659/752)
of the cases in the LCDB reference group.  We posit that the accuracy rate is similar for asteroids that are not in the LCDB reference group.
The best-fit, secondary (mirrored about 3.7831 rot/day), and tertiary (mirrored about 7.5662 rot/day) solutions are accurate in 55\%, 27\%, and 6\% of the cases,  respectively.

Inaccurate solutions that are not mirrors of correct values are visible outside of the X pattern on Figure~\ref{fig-pp-x-plot}.  Figure~\ref{fig-failure} illustrates a failed solution.

A number of solutions deemed to be inaccurate cluster near the accurate solutions along the blue diagonal, especially at low frequencies.  This behavior indicates that our 5\% criterion is a conservative metric of accuracy, and that additional solutions are in fact close to the correct value.

\movetabledown=32mm
\begin{rotatetable}
\begin{deluxetable}{rrrrrrrrrrrrrrrrr}
    \tablecaption{Spin period solutions for 1929 asteroids. \label{tab-results}}
\tablehead{
\colhead{Object} & \colhead{$D$} & \colhead{span} & \colhead{idx} & \colhead{$A$} & \colhead{$P$} & \colhead{$\sigma_P$} & \colhead{$f$} & \colhead{$f_{\xleftrightarrow{1}}$} & \colhead{$P_{\xleftrightarrow{1}}$} & \colhead{$f_{\xleftrightarrow{2}}$} & \colhead{$P_{\xleftrightarrow{2}}$} & \colhead{$P_{\rm LCDB}$} & \colhead{$A_{\rm min}$} & \colhead{$A_{\rm max}$} & \colhead{$U$} & \colhead{flag} \\  
\colhead{      } & \colhead{(km)} & \colhead{(hr)} & \colhead{       } & \colhead{(mag)} & \colhead{(hr)} & \colhead{(hr)}       & \colhead{(rot/d)} & \colhead{(rot/d)} & \colhead{(hr)} &  \colhead{(rot/d)} & \colhead{(hr)}     & \colhead{(hr)}         & \colhead{(mag)}  & \colhead{(mag)}  & \colhead{}  & \colhead{}  
  }

\startdata
131 & 30.6 & 109.5 & 1 & 0.358 & 5.1919 & 0.019 & 4.6226 & 2.9436 & 8.1532 & 10.5098 & 2.2836 & 5.1812 & 0.080 & 0.320 & 3 & 1 \\
155 & 44.6 & 93.7 & 1 & 0.234 & 5.2919 & 0.016 & 4.5352 & 3.0310 & 7.9183 & 10.5972 & 2.2648 & 7.9597 & 0.110 & 0.460 & 3 & 2 \\
170 & 35.4 & 181.0 & 1 & 0.360 & 13.0199 & 0.245 & 1.8433 & 5.7229 & 4.1937 & 9.4095 & 2.5506 & 13.1200 & 0.130 & 0.300 & 3 & 1 \\
180 & 24.8 & 96.9 & 1 & 0.493 & 23.6223 & 0.486 & 1.0160 & 6.5502 & 3.6640 & 8.5822 & 2.7965 & 23.8660 & 0.420 & 0.600 & 3 & 1 \\
183 & 30.7 & 90.5 & 1 & 0.395 & 11.7553 & 0.094 & 2.0416 & 5.5246 & 4.3442 & 9.6078 & 2.4980 & 11.7700 & 0.200 & 0.310 & 3 & 1 \\
183 & 30.8 & 106.4 & 2 & 0.431 & 11.8183 & 0.114 & 2.0308 & 5.5354 & 4.3357 & 9.5970 & 2.5008 & 11.7700 & 0.200 & 0.310 & 3 & 1 \\
...    & ...  & ...   &...&  ...    &  ...  & ...    & ...    & ...     & ...     & ... & ...\\
244 & 10.9 & 20.6 & 1 & 1.152 & 3.2773 & 0.085 & 7.3230 & 0.2432 & 98.6986 & 7.8094 & 3.0732 & 129.5100 & 0.800 & 0.820 & 3- & 0 \\
296       &      9.0  &  141.3 & 1 & 0.390  &  4.5431  &   0.014 & 5.2828 & 2.2834  & 10.5105  & 9.8496   & 2.4366  &  4.5385 & 0.380 & 0.700 & 3  & 1  \\
2715      &    14.0   &  36.5 & 1  &0.584  & 33.1949 &    2.054 & 0.7230 & 6.8432  &  3.5071  & 8.2892  &  2.8953 & 33.6200 & 0.540 & 0.550 & 3- & 1  \\
2812      &     6.0   &  30.2 & 1  & 0.370  &  7.7367  &   0.290 & 3.1021 & 4.4641  &  5.3762  & 10.6683  &  2.2497  &  7.7000    &  &  0.250 & 3  & 1  \\
4715     &     71.4  &   87.3 & 1 & 0.487  &  4.9620 &    0.010 & 4.8367 & 2.7295   & 8.7930  & 10.2957  &  2.3311  &  8.8129 & 0.390 & 0.540 & 3  & 2   \\
6485     &       3.7 &   123.8 & 1 & 0.837 &   3.3112 &   0.035 & 7.2481 & 0.3181 &  75.4453  &  7.8843  &  3.0440  & 75.5600 & 1.000 & 1.130 & 3- & 2  \\
10721     &      4.7   &  42.9 & 1 & 0.852  &  3.2721  &   0.032 & 7.3347 & 0.2315  &  103.6603  & 7.7977  &  3.0778  &  3.0675   &    &  0.270 & 3  & 3  \\
...    & ...  & ...   &...&  ...    &  ...  & ...    & ...    & ...     & ...     & ... & ...\\
256155 & 3.1 & 134.9 & 1 & 0.550 & 15.6910 & 0.118 & 1.5295 & 6.0367 & 3.9757 & 9.0957 & 2.6386 &  &  &  &  &  \\
307840 & 6.8 & 335.1 & 1 & 0.368 & 16.0317 & 0.151 & 1.4970 & 6.0692 & 3.9544 & 9.0632 & 2.6481 &  &  &  &  &  \\
318081 & 5.9 & 42.9 & 1 & 0.410 & 3.6023 & 0.029 & 6.6623 & 0.9039 & 26.5523 & 8.4701 & 2.8335 &  &  &  &  &  \\
366774 & 0.9 & 58.8 & 1 & 0.327 & 3.4972 & 0.036 & 6.8625 & 0.7037 & 34.1075 & 8.2699 & 2.9021 &  &  &  &  &  \\
386720 & 1.0 & 141.3 & 1 & 0.806 & 4.6034 & 0.015 & 5.2136 & 2.3526 & 10.2013 & 9.9188 & 2.4196 &  &  &  &  &  \\
\enddata 
\tablecomments{For each object, we show the WISE-derived diameter $D$ (km) \citep{myhr22}, observation span (hr), index of the cluster of observations analyzed, the best-fit period $P$ (hr), the peak-to-peak magnitude variation of the fitted lightcurve $A$
  for the primary (best-fit) period solution, the primary period uncertainty $\sigma_P$ (hr), the best-fit rotational frequency $f$ (rot/d), the first alternate rotational frequency
  $f_{\xleftrightarrow{1}}$ (rot/d) found by folding the best-fit frequency about 3.7831 rot/day, 
  the first alternate period
  $P_{\xleftrightarrow{1}}$ (hr), the second alternate rotational frequency
  $f_{\xleftrightarrow{2}}$ (rot/d) found by folding the best-fit frequency about 7.5662 rot/day, 
  the second alternate period
  $P_{\xleftrightarrow{2}}$
  (hr),
  the LCDB period $P_{\rm LCDB}$ (hr), the LCDB's minimum amplitude $A_{\rm min}$ if a range was provided, the LCDB's amplitude or maximum amplitude $A_{\rm max}$ if a range was provided, the corresponding quality code $U$, and a flag indicating agreement for objects in the LCDB Reference Group.
The flag is 1 if the best-fit period matches $P_{\rm LCDB}$ within 5\%, 2 if the first mirror period matches, 3 if the second mirror period matches, and 0 if none of the three periods match $P_{\rm LCDB}$. 
Spin period solutions for asteroids shown in Figures \ref{fig-suc-mode-1}, \ref{fig-suc-mode-2}, \ref{fig-suc-mode-3}, \ref{fig-mirror-mode-1}, \ref{fig-mirror-mode-2}, \ref{fig-mirror-mode-3}, and \ref{fig-failure} are included.
(This table is available in its entirety in \href{https://ucla.box.com/s/49z4fr6pdaalfssg7qhngb30vrwjl9f0}{machine-readable} and \href{https://ucla.box.com/s/5muceha0gyj7ynsnysfwjt4uxgln7c8t}{CSV} forms in the online journal. A portion is shown here for guidance regarding its form and content.)}
\end{deluxetable}
\end{rotatetable}

\newpage

\begin{figure}[hbt]
  \begin{center}
         \includegraphics[width=3.5in]{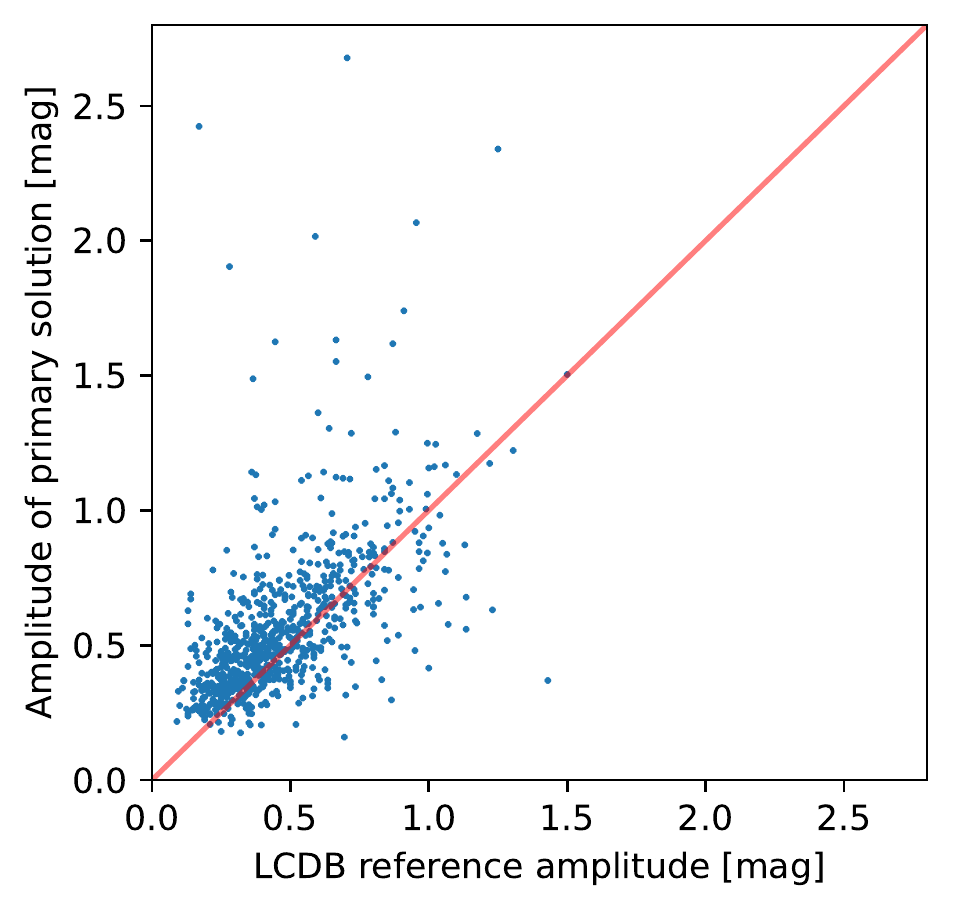}
     \caption{Scatter plot of the lightcurve amplitudes of our primary solutions for the 752 objects in the LCDB reference group and corresponding LCDB amplitudes.
     }
\label{fig-amplitudes}
\end{center}
\end{figure}

We also validated the amplitudes of our primary lightcurve solutions by comparing them to the amplitudes in the LCDB reference group.
  For this comparison, we used the arithmetic average of the LCDB's AMIN and AMAX values when more than one amplitude value was provided in the LCDB.
  By and large there is good agreement, although our values appear to have an overall bias of approximately 0.1 mag (Figure~\ref{fig-amplitudes}).
  In the LCDB reference group, 8 out of 752 (1\%) solutions have an amplitude that exceeds the LCDB amplitude by 1 mag, and 22 additional solutions (3\%) have an amplitude that exceeds the LCDB amplitude by 0.5 mag.  
  We observed that large rotational phase gaps in the timeseries can occasionally result in excessive amplitudes.  The worst-case examples is asteroid 37152, which happens to have a spin period almost exactly equal to three times the WISE cadence.  The second worst-case example is asteroid 4825, which has 
  rotational phase gaps of up to $\sim$0.4.  Our best-fit amplitude (2.678 mag) exceeds the LCDB published value (0.705 mag) by $\sim$2 mag.

\subsection{Lomb-Scargle Pipeline}
\label{res-ls}

As in our default pipeline, we generated three candidate period solutions for each fitted lightcurve (primary: LS solution, secondary: mirror across 3.7831 rot/day, tertiary: mirror across 7.5662 rot/day).  We compared the accuracy of the LS pipeline to our default pipeline by computing the number of solutions that are within 5\% of the high-quality (3/3-) LCDB solutions (Table \ref{tab-LS_results}).

\begin{table}[h]
  \begin{center}
    \begin{tabular}{lrrr}

    Accuracy flag & LS first order & LS second order & Default pipeline \\
    \hline
 0 (inaccurate) &  91 (12\%)  & 518 (52\%) &  93 (12\%) \\
 1 (primary)    &  320 (43\%) & 316 (32\%) & 414 (55\%) \\
 2 (secondary)  &  286 (39\%) & 105 (11\%) & 203 (27\%) \\
 3 (tertiary)   &  33 (5\%)   & 56 (6\%)   & 42 (6\%) \\
 Aggregate accuracy & 88\%   & 48\%      &  88\% \\
 Number of reference solutions & 730 & 995 & 752  \\

  \end{tabular}
    \caption{ Number of accurate solutions with the Lomb-Scargle and default pipelines.
      }
    \label{tab-LS_results}
\end{center}  
  \end{table}

The first-order LS solutions have an aggregate accuracy comparable to the default pipeline solutions. However, the primary LS solutions were accurate only in 43\% of the cases, compared to 55\% in the default pipeline. 
The second-order LS solutions have a lower accuracy rate than the default pipeline solutions, both for primary solutions and in aggregate. The lower performance of the LS algorithm demonstrates the importance of the
iterative algorithm and post-fit filters in our default, \citet{wasz15}-inspired pipeline.

\FloatBarrier
\section{Discussion}
\label{sec-discussion}

\begin{figure}[hbt]
 \begin{center}
     \includegraphics[width=4.5in]{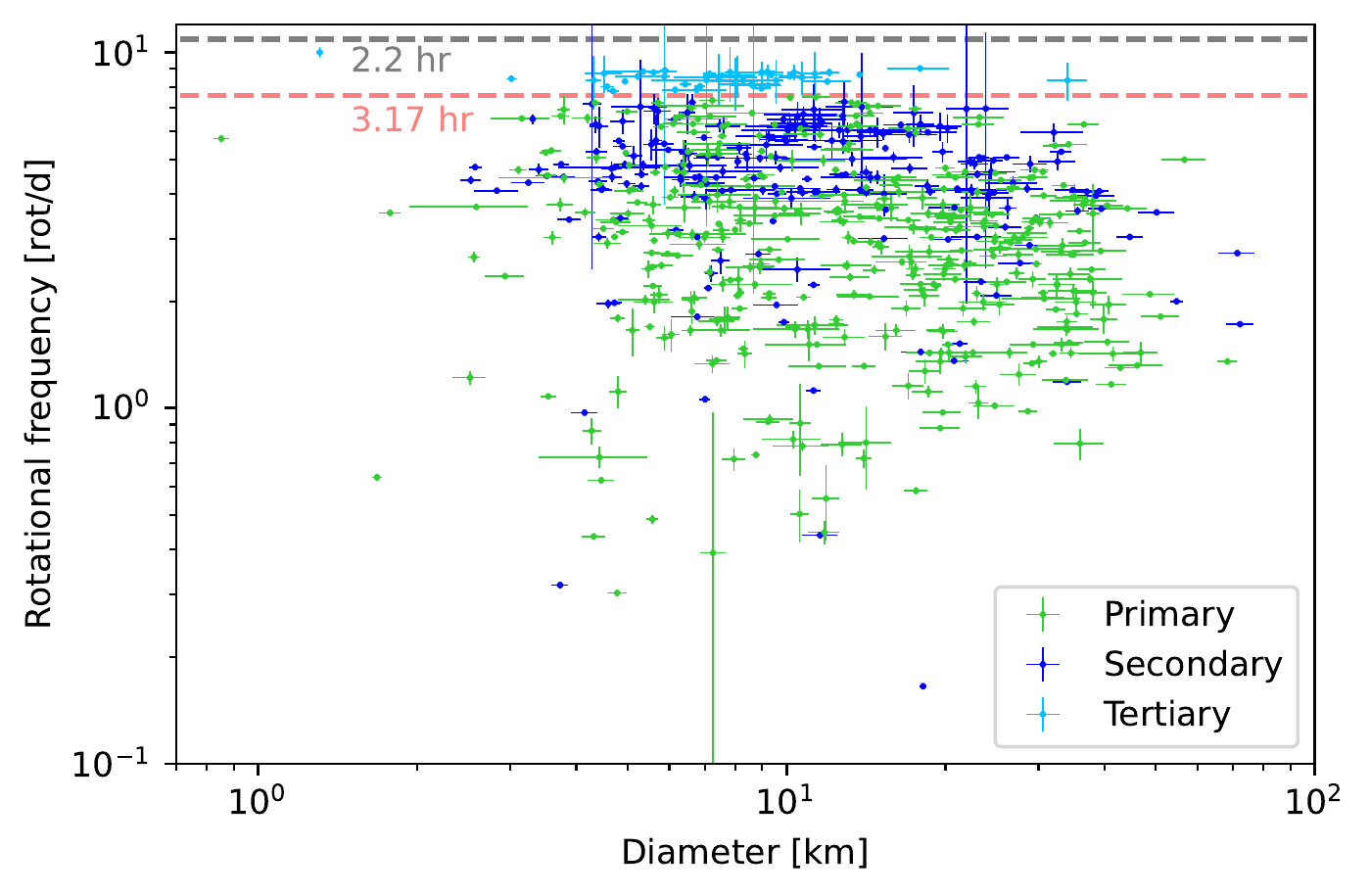}
          \caption{Rotational frequency vs.\ diameter diagram for the 659 light curves in the LCDB reference group where one of our spin period solutions matched the LCDB to 5\%.  Only the spin period solutions that match the LCDB value are plotted, with color-coding indicating our primary, secondary, or tertiary solutions. The red dashed line at $P=$ 3.17 hr corresponds to the upper range of trial frequencies explored in the fitting process. The grey dashed line at $P=$ 2.2 hr illustrates the spin barrier. Diameters are from  \citet{myhr22}.}
\label{fig-period_diam_accurate_fit_period}
\end{center}
\end{figure}

\citet{prav02} reviewed the rotation periods of asteroids as a function of diameter and found that the spin distribution
is Maxwellian for large asteroids ($>$40 km) and strongly non-Maxwellian for smaller asteroids, with an excess of both slowly rotating and rapidly rotating asteroids.
The WISE data set can potentially inform these studies because it can yield estimates of both diameter and spin period (Figure~\ref{fig-period_diam_accurate_fit_period}).
However, the value of our estimates in this context is diminished because of obvious selection effects with respect to spin period.

\begin{figure}[hbt]
  \begin{center}
     \includegraphics[width=4.5in]{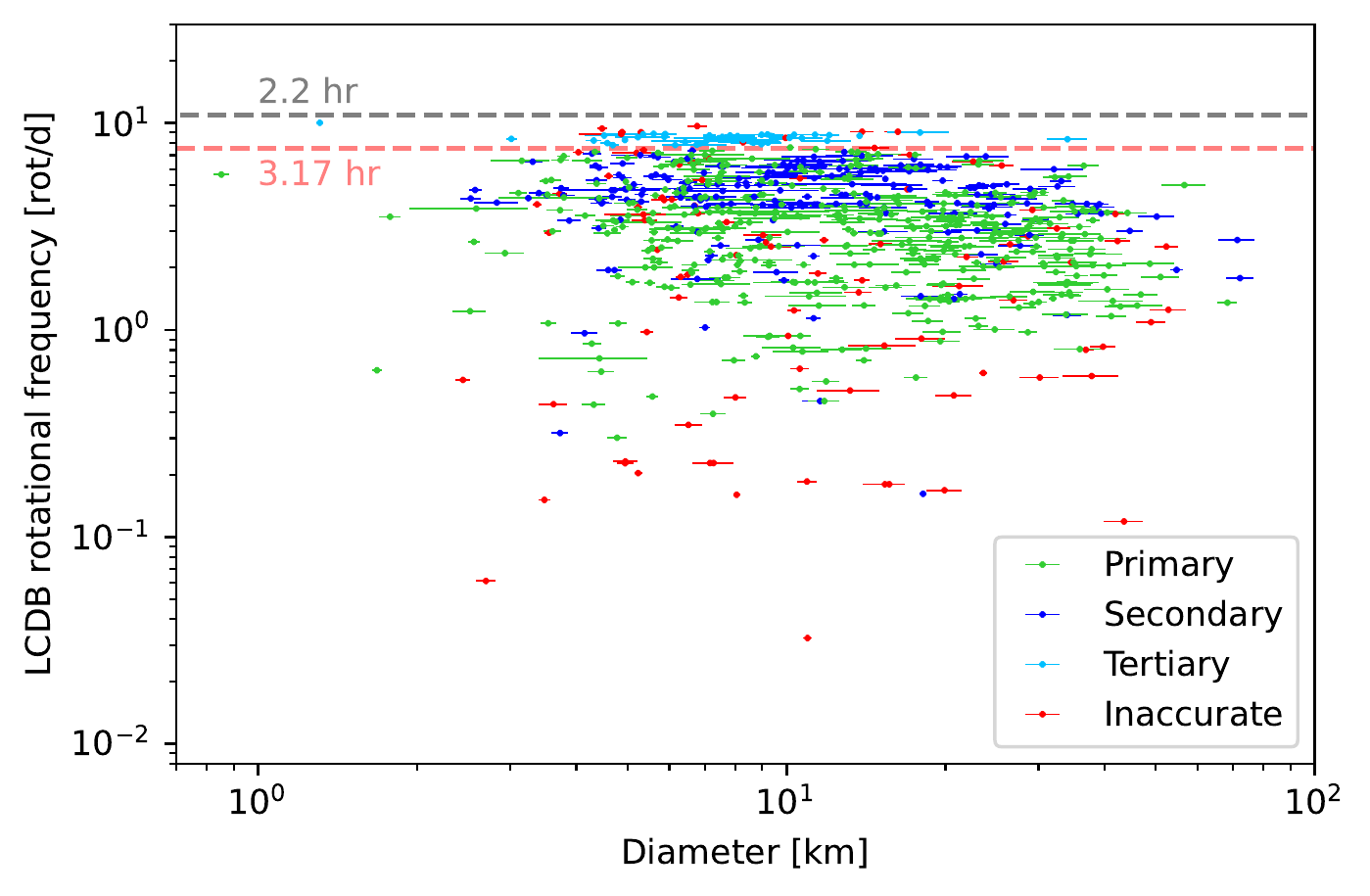}
          \caption{LCDB rotational frequency vs.\ diameter \citep{myhr22} for the 752 lightcurves in the LCDB Reference Group. Our spin period solutions are color-coded according to four possible outcomes: primary solution is accurate, secondary solution is accurate, tertiary solution is accurate, or none are accurate.  The red dashed line at $P=$ 3.17 hr corresponds to the upper range of trial frequencies explored in the fitting process. The grey dashed line at $P=$ 2.2 hr illustrates the spin barrier.  Diameters are from  \citet{myhr22}.}
\label{fig-period_diam_lcdb_ref_period}
\end{center}
\end{figure}

We can evaluate the accuracy of our method as function of asteroid diameter and LCDB period (Figure \ref{fig-period_diam_lcdb_ref_period}).
Diameter does not have an apparent influence on the accuracy of our method.
However, spin period does affect our ability to recover a correct solution with the WISE data.  We found that the spin periods of both slowly rotating ($P > \sim$100 hr) and rapidly rotating  ($P < \sim$2.5 hr) asteroids are generally not recoverable.

\begin{figure}[hbt]
  \begin{center}
     \includegraphics[width=4.5in]{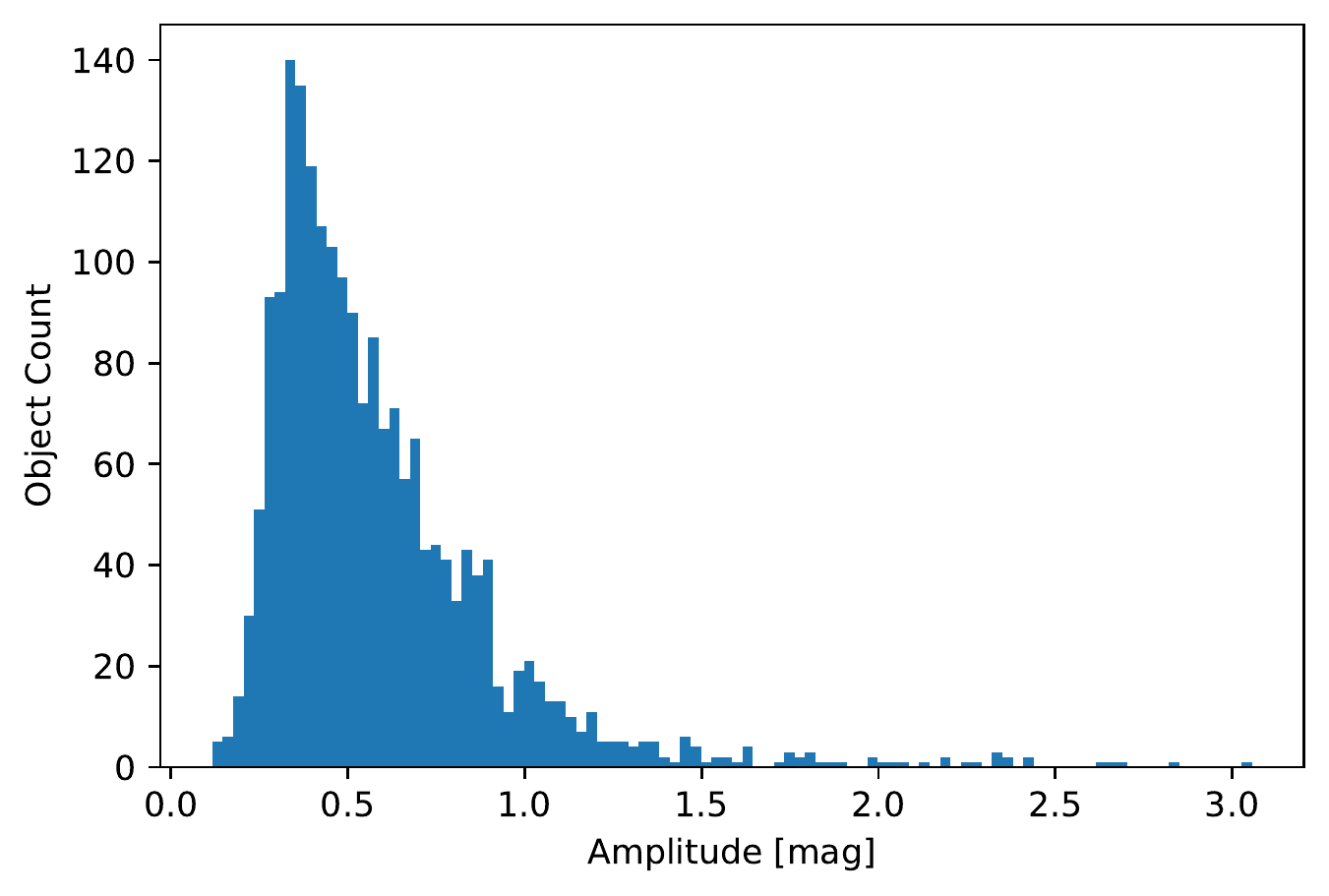}
          \caption{Histogram of peak-to-peak lightcurve amplitudes.}
\label{fig-amp-histogram}
\end{center}
\end{figure}

The distribution of lightcurve amplitudes in our data set is instructive (Figure~\ref{fig-amp-histogram}).  Our distribution underestimates the proportion of low-amplitude lightcurves, because our sample selection enforced a pre-fit filter, which required a magnitude variation of at least 0.3 mag in the observed fluxes.
The bias against low-amplitude lightcurves is not limited to our work.
Likewise, one of our post-fit filters eliminated some high-amplitude lightcurve solutions, but it did so only when the data themselves did not exhibit a large magnitude variation.
We found that 50\%, 8.8\%, 2.2\%, 1\%, and 0.25\% of solutions have amplitudes larger than 0.5 mag, 1 mag, 1.5 mag, 2 mag, and 2.5 mag, respectively.

\added{Aliasing in our solutions was exacerbated by the quasi-uniform cadence of WISE observations.  Although recovery from aliasing is relatively straightforward, it is possible to reduce the presence of aliased solutions with non-uniform sampling.  To illustrate this point, we generated 100 sequences of simulated observations of asteroid 4715 (Figure~\ref{fig-mirror-mode-1}) with $\sim$uniform (WISE) sampling and 100 sequences with non-uniform sampling, all with the same number of observations over the same time interval.  All simulated observations were generated with a sixth-order Fourier series and polluted with random Gaussian noise ($\sigma$=0.05 mag).  Among the 100 uniformly sampled sequences, 89 yielded the correct solution and 11 yielded a (recoverable) aliased solution mirrored around the f = 3.7831 rot/day axis.  All non-uniformly sampled solutions yielded the correct solution.  For this reason, ground-based observers may wish to intentionally deviate from a strictly regular cadence.}

\section{Conclusions}
\label{sec-conclusions}
We devised a procedure similar to that of \citet{wasz15} that enables the determination of thousands of asteroid spin periods from WISE data.  Despite WISE's suboptimal observation and cadence for asteroid spin measurements,
one of our
solutions is accurate 88\% of the time when compared to a high-quality control group of 752 spin periods.  We obtained primary, secondary, and tertiary spin period estimates for 2008 observation clusters representing 1929 unique asteroids.  Among those, 1205 asteroids do not currently have a high-quality spin period estimate.  Our
primary, secondary, or tertiary 
solutions for
over a thousand asteroids are expected to be accurate at the 5\% level or better and can greatly facilitate shape or thermal modeling work.
In addition, they may reveal objects with unusual lightcurves worthy of more thorough observations.

\acknowledgments

We thank Alan Harris (California) and an anonymous reviewer for insightful and constructive reviews.
We thank Robert D. Stephens for providing Fourier coefficients for the lightcurve of 4715.
A.L. was funded in part by the Joe and Andrea Straus Endowment for Undergrad Opportunity
and the Donald Carlisle Undergrad Research Endowed Fund.
E.W. was funded in part by the Nathan P. Myhrvold Graduate Fellowship.
A.L. thanks Breann Sitarski for her helpful feedback on this work and the related research presentations, as well as her mentorship throughout his undergraduate studies.
This publication makes use of data products from the Wide-field Infrared Survey Explorer, which is a joint project of the University of California, Los Angeles, and the Jet Propulsion Laboratory/California Institute of Technology, funded by the National Aeronautics and Space Administration.

\software{
NumPy \citep{numpy},
SciPy \citep{scipy},
pandas \citep{pandas},
Matplotlib \citep{mpl}
}

\appendix

\section{Aliasing}
\label{app-aliasing}
        
The observation cadence, i.e., the series of time intervals between consecutive photometric measurements, is an important factor in the analysis of asteroid spin periods.  If the sampling does not yield at least two points per cycle, aliasing occurs.
Aliasing often produces
replicas of the peak corresponding to the correct period in the power spectrum,
precluding a unique period determination.  These peaks may exhibit some asymmetry in the misfit values displayed in the periodogram, and the correct period is not guaranteed to have the best-fit value.
In the case of asteroid spin determinations from WISE, the problems associated with aliasing and asymmetry are
amplified because of
the sparsity of data points, short observation spans, and measurement noise.

We expect signals with frequencies $>$ \added{7.5662} rot/day to alias and fold about the \added{7.5662} rot/day axis.  We also expect folding about the \added{3.7831} rot/day axis due to presence of sampling intervals at 3.17 hr.  While some mirroring behavior is obvious, it is not always possible to detect which axis was involved in the folding behavior.
We took several actions to maximize the reliability in the recovered period solutions.

First, we limited the range of scanned frequencies to the Nyquist rate corresponding to the 1.59 hr cadence (Section~\ref{sec-fit}). 
A drawback of this approach is that it initially prevents recovery of the spin periods of asteroids that rotate faster than f = \added{7.5662} rot/day, 
which amount to approximately 17\% of asteroids in our sample.  However, a fraction of these solutions are recovered because we provide the solution that is mirrored about the f = \added{7.5662} rot/day axis.

Second, we eliminated time series where the minimum cadence is 3.17 hr or above (Section \ref{sec-data-prefilters}) because these data are most prone to aliasing.  For the same reason, we preferred the W4 data over the W3 data (Section~\ref{sec-band-selection}).
Third, we attempted to reject aliased solutions based on domain knowledge.
We required double-peaked folded lightcurves with a
peak ratio of two or larger (Section~\ref{sec-fit}) to encourage a second-order dominant Fourier series.

\bibliographystyle{aasjournal}
\bibliography{wise_lc}

\end{document}